\documentclass[a4paper,11pt]{article}
\usepackage{paperStylefinale} 
\usepackage{paperCommands} 
\usepackage{paperCommandsDavide}
\newenvironment{dedication}
        {\vspace{6ex}\begin{quotation}\begin{center}\begin{em}}
        {\par\end{em}\end{center}\end{quotation}}
\usepackage{pdfsync}
\hypersetup{
     hypertexnames=false,%
     debug=false,%
     colorlinks=true,%
     linkcolor=black,%
     citecolor=black,%
     urlcolor=black,%
     bookmarksopen=true,%
     bookmarksopenlevel=1,%
     bookmarksnumbered=true,%
     pdfinfo={%
          Title={},
          Subject={},%
          Author={Roberta Springhetti, Gabriel Rossetto, Davide Bigoni} },
     pdflang={English},%
     pdfpagemode={UseOutlines},
     pdfstartview={Fit},%
     pdfdisplaydoctitle=true,%
     breaklinks=true%
}
\ifdraft{
     \PassOptionsToPackage{draft}{bookmark}
     \PassOptionsToPackage{draft}{hyperref}
     \usepackage[inline]{showlabels}%
     
}{}
\title{
Buckling of thin-walled cylinders \\from three dimensional nonlinear elasticity
}
\author[a]{Roberta Springhetti\footnote{E-mail:\,roberta.springhetti@unitn.it;}}
\author[a]{Gabriel Rossetto\footnote{E-mail:\,rossetto.gabriel@gmail.com;}}
\author[a]{Davide Bigoni\footnote{Corresponding author:\,e-mail:\,bigoni@ing.unitn.it; phone:\,+39\,0461\,282507.}}
\affil[a]{DICAM, University of Trento via Mesiano 77, Trento, Italy}
\begin{document}
\date{}
\maketitle
%
%
\vspace*{-2.1cm}
\begin{dedication}
\begin{center}
{paper accepted for publication in the Journal of Elasticity}\\
\vspace*{1cm}
{\it Dedicated to Professor Roger L. Fosdick}
\end{center}
\end{dedication}
\begin{abstract}
The famous bifurcation analysis performed by \Fluegge on compressed thin-walled cylinders is based on a series of simplifying assumptions, which allow to obtain the bifurcation landscape, together with explicit expressions for limit behaviours: surface instability, wrinkling, and Euler rod buckling. 
The most severe assumption introduced by \Fluegge is the use of an incremental constitutive equation, which does not follow from any nonlinear hyperelastic constitutive law. This is a strong limitation for the applicability of the theory, which becomes questionable when is utilized for a material characterized by a different constitutive equation, such as for instance a Mooney-Rivlin material. 
We re-derive the entire \Fluegge's formulation, thus obtaining a framework where any constitutive equation fits. The use of two different nonlinear hyperelastic constitutive equations, referred to compressible materials, leads to incremental equations, which reduce to those derived by \Fluegge under suitable simplifications. His results are confirmed, together with all the limit equations, now rigorously obtained, and his theory is extended. This extension of the theory of buckling of thin shells 
allows for computationally efficient determination of bifurcation landscapes for nonlinear constitutive laws, which may for instance be used to model biomechanics of arteries, or soft pneumatic robot arms. 
\end{abstract}
%
%
\noindent Keywords: thin shells, nonlinear elasticity, \FvK{}'s theory, \Fluegge{}'s buckling load
%
%
%
\section{Introduction}
	\label{Sec-Introduction}
Buckling of thin-walled cylinders subject to axial thrust represents one of the most famous problems in mechanics and a fascinating question in bifurcation theory. In fact it is well-known\footnote{Reviews on this topic are reported in many books and articles on structural stability, see for instance \textcites{Budiansky1974}{Calladine1983}.}
that the critical load for buckling (calculated in a linearized context by \textcites{Lorenz1908}{Timoshenko1910}{Southwell1914}{vonMises1914}{Flugge1932}{Donnell1933} and elegantly reported by \textcites{Flugge1962} and \textcite{Yamaki1984}) provides only an overestimation of the carrying capacity which can 
experimentally 
be  measured on real cylinders. This overestimation was explained in terms of post-critical behavior in a number of celebrated works (among which, \textcites{Karman1941}{Koiter1945}{Hutchinson1965}{Hutchinson1970Postbuckling}{Tsien1942}). 
Fifty years later, the mechanics of thin shells remains a prosperous research topic (\textcites{Lee2016}{Jimenez2017}{Elishakoff2014}{Ning2016}), also embracing recent applications to nanotubes (\textcites{Waters2005}) and soft materials, the latter developed as a key to understand biological systems, (\textcites{Lin2018}{Steele1999}) or towards mechanical applications, (\textcites{Leclerc2020}{Shim2012}). 

The bifurcation analysis performed by \Fluegge is based on a series of approximations, among which, the  
incremental constitutive equations do 
{\it not follow from a finite strain formulation of any hyperelastic material}.
In particular, it is shown that the equations relate, through a fourth-order isotropic elastic tensor, the Oldroyd increment of the Kirchhoff stress to the incremental Eulerian strain. These equations, involving Lam\'e moduli $\lambda$ and $\mu$ are certainly valuable in an approximate sense, but how this approximation may be tied to 
a rigorous theory of nonlinear elasticity remains unknown. 

The focus of the present article is the incremental\footnote{
An incremental analysis is, in other words, \lq linearized', so that the post-critical behavior is not considered.
}
bifurcation analysis of an axially-loaded thin-walled cylinder, characterized by rigorously-determined, nonlinear hyperelastic constitutive equations. 
Our analysis generalizes and rationalizes the famous derivation performed by \Fluegge not only from the point of view of the constitutive equations, 
but also because it allows to either rigorously prove, or clearly elucidate other assumptions. 
In particular, the \Fluegge derivation is based on the smallness assumption for the thickness of the cylinder wall. 
This represents an approximation on the one hand and a simplification 
on the other. There are only three alternatives to circumvent this approximation, namely: 
(i.) a numerical approach (for instance through a finite element code),
but numerical solutions are approximated and far from providing the deep insight and the generality intrinsic to a theoretical determination; 
(ii.) a direct approach from three-dimensional incremental elasticity
(as for instance pursued by \textcites{Wilkes1955}, \textcite{HAUGHTON1979489}, \textcites{Bigoni2001}{Chau1995}), but the numerical solution of the bifurcation condition involved in this technique becomes awkward in the thin-walled limit; 
(iii.) a reduction (if possible) of the nonlinear elastic constitutive laws to a small-strain version, based on Lam\'e constants to be used in the \Fluegge equations, but in doing so, an unknown approximation is introduced.

The three above-mentioned alternatives are abandoned in this article (except for the `direct approach' that will be used to validate the obtained results), in favor of a re-derivation of the buckling of a thin-walled cylinder, pursued from a different perspective.
First, the incremental equilibrium equations are rigorously derived in terms of mean quantities, represented by generalized stresses (holding true regardless of the thickness of the cylinder), through a generalization of the approach introduced by \textcite{Biot1965} for rectangular plates. The incremental kinematics is postulated as a deduction from the deformation of a two-dimensional surface, again in analogy with the incremental kinematics of a plate. Our treatment of thin-walled cylinders {\it allows the use of every nonlinear constitutive law}. 
In particular, two different nonlinear elastic constitutive equations describing compressible neo-Hookean materials (Pence and Gou, 2015) are rigorously used. While the linearized kinematics adopted coincides with that used by \Fluegge, the incremental equilibrium equations derived in this article are different from \Fluegge's corresponding equations, but are shown to reduce to the latter by invoking smallness of the cylinder wall thickness.
The equations obtained for the incremental deformation of prestressed thin cylindrical shells (Sections \ref{Sec-Field-Equations}--\ref{Sec-Constitutive-Equations}) are general and can be used for different purposes, so that the ensuing bifurcation analysis represents only an example of application, while other problems can be pursued, such as for instance, the torsional buckling. When compared  (Section \ref{bif-cyl}), the bifurcation landscapes obtained from our formulation and that given by \Fluegge are shown to be almost coincident and perfectly consistent with results obtained through the `direct approach', where the fully three-dimensional problem is solved (which is also a new result presented here in Section \ref{Sec-3D-Solution}). Finally, the following formulas are rigorously obtained as limits of our approach: 
(i.) the surface instability, in the 
short longitudinal wavelength limit; 
(ii.) the wrinkling, occurring as axial buckling of a mildly long cylindrical shell, characterized by the well-known formula obtained by  \Fluegge, (iii.) the Euler rod buckling for a long cylindrical shell (Section \ref{limiti}). 

The re-derivation of the \Fluegge formulation within a three-dimensional finite elasticity context, including the calculations of the bifurcation loads and the determination of the famous formula for buckling of a mildly long cylindrical shell, is important from two different perspectives. 
First, the validity of the \Fluegge theory, considered a  reference in the field, is now confirmed. 
Second, the new derivation is applicable to soft materials,  characterized in the framework of nonlinear elasticity
by constitutive equations different from those used by \Fluegge. Therefore, the determination of the buckling stress is now possible for a cylindrical shell made up of an Ogden or a neo-Hookean compressible elastic material \parencites{Levinson1971}{Ogden1972a}, or for an artery obeying the \textcite{Holzapfel2000a} constitutive law.
%
%
\section{Incremental field equations in terms of generalized stresses}
     \label{Sec-Field-Equations}
The undeformed stress-free configuration is described by means of cylindrical coordinates  $(r_0, \theta_0, z_0)$, being the $z_0$-axis aligned parallel to the axis of revolution of the shell. Along its fundamental path before bifurcation, the shell is assumed to undergo a homogeneous, axisymmetric compression in its longitudinal direction $z_0$, preserving the circular cylindrical geometry. A uniaxial stress is generated in the form
\begin{equation} \label{Eq-Def-tensK}
     \tensK = \Kz \tensG\comma
\end{equation}
where $\tensK = J \tensT$ represents the \Kirchhoff stress tensor, with $J = \det{\tensF}$, being $\tensF$  the deformation gradient and $\tensT$ the \Cauchy stress, while $\tensG = \ez \otimes \ez$ ($\ez$ is the unit vector singling out the $z$-axis). 
The current configuration is described through coordinates $(r, \theta, z)$ by means of the principal stretches $\{\lambdar, \lambdatheta, \lambdaz\}$ as
\begin{equation*}
     r = \lambdar \, r_0, \quad \theta =  \theta_0, \quad z = \lambdaz \, z_0 \comma
\end{equation*}
with $\lambdar=\lambdatheta$ following from axial symmetry.
Therefore, the deformation gradient and the left \CG deformation tensor read as \mbox{$\tensF = \mathrm{diag}\{\lambdar; \, \lambdatheta; \, \lambdaz\}$} and \mbox{$\tensB = \tensF \tensFT = \mathrm{diag}\{\lambdar^2; \lambdatheta^2; \lambdaz^2\}$}, respectively.

The incremental equilibrium equations are derived, governing the bifurcations of a cylindrical shell of current length $l$, external radius $\rext$ and internal radius $\rint$.
The cylinder, whose thickness is denoted by $t = \rext - \rint$,  is not assumed to be thin for the moment, therefore all results presented in this Section are rigorous in terms of mean values of the incremental field quantities.
The geometrical descriptors adopted here are the mid-radius \mbox{$a = (\rext + \rint) /2$}, defining the `mid-surface' of the shell, and the so-called reduced radius $\rred = r - a$.
A standard notation is used, where bold capital and lower case letters denote tensors and vectors, respectively.
 
Adopting the relative Lagrangean description, with the current configuration assumed as reference, such that $\tensF = \tensI$, and neglecting the body forces, the incremental equilibrium of a pre-stressed solid is expressed through $\tensdotS$, the increment of the first \PK stress tensor $\tensS$, as
\begin{equation}
     \label{Eq-Incremental-Equilibrium}\pcomma
    \div \tensdotS = \vec{0}.
\end{equation}
The cylindrical shell is subject to traction-free surface boundary conditions on its lateral surface, so that 
\begin{equation}\label{EqBC-dotS}\ppoint
     \tensComponent{\dotS}{i}{r} = 0 \,\,\,\,\, \mbox{as}\,\,\,\,  \rred = \pm t/2 \quad( i = r, \theta, z).
\end{equation}

The increment of the \Kirchhoff stress $\tensK$ can be related to $\tensdotS$ through equation $\tensS = \tensK \tensFmt$, namely
\begin{equation}
     \label{Def-dotS}
		\pcomma
     \tensdotS = (\tensdotK - \tensK \tensL^\T) \, \tensFmt \comma
\end{equation}
where $\tensL = \grad{\vecv}$ is the gradient of the incremental displacement field $\vecv$. In a relative Lagrangean description, equation~\eqref{Def-dotS} becomes 
\noindent
\begin{equation}
     \label{Def-dotSUpdateLagrangean}\ppoint
     \tensdotS =\left(\tr{\tensL}\right)\tensT  + \tensdotT - \tensT \tensL^\T\!\point
\end{equation}

Introducing the uniaxial pre-stress,  \EqRef{Eq-Def-tensK}, into \EqRef{Def-dotSUpdateLagrangean}, the following relations between the 
components of the incremental first Piola-Kirchhoff stress tensor $\tensdotS$ are derived:
\begin{equation}\label{Eq-dotS-Component-Transformation}\ppoint
	\begin{gathered}
		\dotSthetar = \dotSrtheta \comma\\ 
		\dotSzr = \dotSrz - \pardz{\vr} \, \Kz \comma\\
		\dotSztheta = \dotSthetaz - \pardz{\vtheta} \, \Kz\point
	\end{gathered}
\end{equation}
%
%
\subsection{Exact formulation}

\subsubsection{Generalized stresses}
	\label{SubSubSec-Generalized-Stresses}
In the shell theory, it is common to introduce the so-called \lq generalized stresses', namely, stress resultants per unit length referred to the mid-surface of the shell. For a cylinder of current uniform wall thickness $t = \lambdar \, t_0$, the following definitions are adopted for the increments of forces and moments:
\begin{equation} \label{Eq-Def-Generalized-FM}
\begin{gathered}
     \Genericderiv{n}_{\,\cdot \theta} = \intrred{\Genericderiv{\mbox{stress}}_{\,\cdot \theta}},%
     \quad \quad\,\,\,\,\,%
     \Genericderiv{n}_{\,\cdot z} = \intrred{\Genericderiv{\mbox{stress}}_{\,\cdot z} \, \left( 1 + \rred/a \right)\! }, \quad\,\,\,\,\,           
\\
     \Genericderiv{m}_{\,\cdot \theta} = - \intrred{\Genericderiv{\mbox{stress}}_{\,\cdot \theta}\, \, \rred},%
     \quad %
     \Genericderiv{m}_{\,\cdot z} = - \intrred{\Genericderiv{\mbox{stress}}_{\,\cdot z} \,\,  \rred \left( 1 + \rred/a \right)\!},         
\end{gathered}
\end{equation}
where the subscript $\cdot$ stands for $r$, $\theta$, or $z$ in turn, while
\lq $\mbox{stress}_{\,\cdot z}$' 
 and  
 \lq $\mbox{stress}_{\,\cdot \theta}$' 
represent 
the $\cdot z$ and the $\cdot \theta$ component of a generic
 Eulerian stress measure. The  superimposed $\star$ identifies a suitable increment, in particular here symbols $\cdot$ and $\circ$ are used to denote material time derivative and \Oldroyd derivative, respectively. The factor $1+\rred/a$ in  $\Genericderiv{n}_{\,\cdot z}$ and $ \Genericderiv{m}_{\,\cdot z}$ is the consequence of the integration over a circular sector.

The following generalized stresses play a role hereafter:
\begin{center}
     \begin{tabular}{l l l l }
     	$\nrtheta$ &radial shear force,          &   $\quad\nrz$ & transverse shear force,                \\
     	$\ntheta$ &hoop force,                    &$\quad\nthetaz$ & circumferential membrane shear force,    \\
     	$\nztheta$ &longitudinal membrane shear force,  &$\quad\nz$  & longitudinal normal  force, \\
        $\mtheta$  &hoop bending moment,               &$\quad\mthetaz$ & longitudinal twisting moment, \\
     	$\mztheta$ &circumferential twisting moment,     &$\quad\mz$ & circumferential bending moment.
     \end{tabular}
\end{center}

%
%
\subsubsection{Incremental equilibrium equations: material formulation}
     \label{SubSec-Equilibrium-Equations-dotS-Exact}
In a polar coordinate system $\{\er, \etheta, \ez\}$, \EqRef{Eq-Incremental-Equilibrium} corresponds to the three scalar equations
\begin{equation}\label{Eq-Indefinite-Incremental-Equilibrium-Cylinder}
     \begin{dcases}
           (a+\rred) \, \left( \pardr{\dotSr}+\pardz{\dotSrz} \right) + \pardtheta{\dotSrtheta} + \dotSr - \dotStheta  = 0 \comma\\%
           (a+\rred) \, \left(\pardr{\dotSthetar} + \pardz{\dotSthetaz} \right) + \pardtheta{\dotStheta} + \dotSrtheta + \dotSthetar   = 0 \comma\\%
           (a+\rred) \, \left( \pardr{\dotSzr} + \pardz{\dotSz} \right)  + \pardtheta{\dotSztheta} + \dotSzr = 0 .%
     \end{dcases}
\end{equation}
Focusing now on  \EqRef{Eq-Indefinite-Incremental-Equilibrium-Cylinder}$_2$, after multiplication by the reduced radius $\rred$, a through-thickness integration yields
\begin{equation} \label{Eq-Example-Derivation-01-01}
	a \intrred{ \left( \pardr{\dotSthetar} + \pardz{\dotSthetaz}\right) \, (1+\rred/a) \, \rred}+
	\intrred{ \pardtheta{\dotStheta} \, \rred } +
	\intrred{ ( \dotSrtheta + \dotSthetar ) \, \rred }  = 0 .
\end{equation}
The derivatives of the generalized moments $\dotmtheta$ and $\dotmthetaz$ according to~Eqs.~\eqref{Eq-Def-Generalized-FM} can easily be  recognized in the above equation, while an integration by parts allows to transform the first term as
\begin{equation*}
     \intrred{ \pardr{\dotSthetar} \, (1+\rred/a) \, \rred } =
     -  \, \intrred{ \dotSthetar  \, \left( 1 + 2 \, \rred/a \right) } +  \, \evalThickness{[ \dotSthetar  \, \left( 1 + \rred/a \right)\, \rred] \,} \comma
\end{equation*}
so that, exploiting \EqRef{Eq-dotS-Component-Transformation}$_{1}$, \EqRef{Eq-Example-Derivation-01-01} becomes
\begin{equation} \label{Eq-Example-Derivation-01-02}
     \pardtheta{\dotmtheta}%
     + a \, \pardz{\dotmthetaz}%
     + a\, \dotnrtheta%
     - \evalThickness{ [ \dotSthetar \, \rred \, \left( a + \rred \right)]\, }%
	= 0 \point%
\end{equation}
The same procedure can be applied to \EqRef{Eq-Indefinite-Incremental-Equilibrium-Cylinder}$_3$ after multiplication by $\rred$ and subsequent integration to generate the next rotational equilibrium equation
\begin{equation} \label{Eq-Example-Derivation-02}
         a \, \pardz{\dotmz} + \pardtheta{\dotmztheta}%
     + a \, \dotnrz %
     - P \, a / t \intrred{ v_{r,z} \, \left( 1 + \rred/a \right) }\,
     -  \evalThickness{[ \dotSzr \, \rred \, \left( a + \rred \right)]\,}%
     = 0 \comma
\end{equation}
where \mbox{$P = \Kz \, t$} represents the pre-stress load per unit length along the mid-circular surface, multiplied by $J$.
\noindent From a mechanical point of view, Eqs.~\eqref{Eq-Example-Derivation-01-02} and \eqref{Eq-Example-Derivation-02} enforce the equilibrium of moments about the $z$- and $\theta$- axes, respectively.

The three translational equilibrium equations for the generalized stresses are obtained in a similar vein, through a direct through-thickness integration of  Eqs.~\eqref{Eq-Indefinite-Incremental-Equilibrium-Cylinder} with an integration by parts 
\begin{equation}\label{Eq-Integral-Incremental-Equilibrium-01}
\begin{dcases}
          \pardtheta{\dotnrtheta}     + a \, \pardz{\dotnrz} - \dotntheta%
         + \evalThickness{[ \dotSr \, \left(a+\rred \right)]\,}= 0\comma \\%
          \pardtheta{\dotntheta} + a \, \pardz{\dotnthetaz} + \dotnrtheta%
          + \evalThickness{[ \dotSthetar \, \left(a+\rred \right)]\, }%
          = 0 \comma \\%
           a \, \pardz{\dotnz} + \pardtheta{\dotnztheta}%
          + \evalThickness{[ \dotSzr \, \left(a+\rred \right)]\, } %
          = 0 \point
     \end{dcases}
\end{equation}

Enforcing the boundary conditions, Eq.~\eqref{EqBC-dotS}, on Eqs.~\eqref{Eq-Example-Derivation-01-02}--\eqref{Eq-Integral-Incremental-Equilibrium-01}, the full system of equilibrium equations is finally obtained 
\begin{equation}\label{Eq-Integral-Incremental-Equilibrium-02}
     \begin{dcases}
         \pardtheta{\dotnrtheta}     + a \, \pardz{\dotnrz} - \dotntheta%
          = 0 \comma \\%
          \pardtheta{\dotntheta} + a \, \pardz{\dotnthetaz} + \dotnrtheta%
          = 0 \comma \\%
           a \, \pardz{\dotnz} + \pardtheta{\dotnztheta}%
          = 0 \comma \\%
          \pardtheta{\dotmtheta} + a \, \pardz{\dotmthetaz} + a \, \dotnrtheta%
          = 0 \comma \\%
          a \, \pardz{\dotmz} + \pardtheta{\dotmztheta} + a \, \dotnrz - P \, a / t \intrred{ v_{r,z} \, \left( 1 + \rred/a \right) }%
          = 0 \point%
     \end{dcases}
\end{equation}

A substitution of \EqRef{Eq-Integral-Incremental-Equilibrium-02}$_4$ and \EqRef{Eq-Integral-Incremental-Equilibrium-02}$_5$ into \EqRef{Eq-Integral-Incremental-Equilibrium-02}$_1$ and \EqRef{Eq-Integral-Incremental-Equilibrium-02}$_2$ allows to remove the shear forces, thus leading to the following equations: 
\begin{equation}\label{Eq-Integral-Incremental-Equilibrium-Final}\ppoint
     \begin{dcases}
         \parddtheta{\dotmtheta} + a \, \pardpard{(\dotmthetaz+\dotmztheta)}{\theta}{z} + a^2 \, \parddz{\dotmz} + 
         a\, \dotntheta - P \, a^2 / t \intrred{ \parddz{\vr} \, \left( 1 + \rred/a \right) } %
          = 0 \comma \\ %
          a \, \pardtheta{\dotntheta} + a^2 \, \pardz{\dotnthetaz} - \pardtheta{\dotmtheta} - a \, \pardz{\dotmthetaz}%
          = 0 \comma \\%
          a \, \pardz{\dotnz} + \pardtheta{\dotnztheta}%
          = 0 \comma \\
          \dotnrtheta = - \, \pardtheta{\dotmtheta}/a - \pardz{\dotmthetaz} \comma \\%
          \dotnrz  = - \, \pardz{\dotmz} - \pardtheta{\dotmztheta}/a + P / t \intrred{ v_{r,z} \, \left( 1 + \rred/a \right) } \point
     \end{dcases}
\end{equation}
%
%
\subsubsection{Incremental equilibrium equations: spatial formulation}
     \label{SubSec-Equilibrium-Equations-OldroydK}
In a relative Lagrangean description the incremental equilibrium equations~\eqref{Eq-Integral-Incremental-Equilibrium-Final} can equivalently be  expressed by means of a new set of generalized stresses, based on the Oldroyd increment \parencite{Oldroyd1950} of the \Kirchhoff stress $\tensK$, namely
\begin{equation}\label{Eq-Def-tensOldroydK}
     \tensOldroydK = \tensdotS - \tensL \tensK  \point
\end{equation}
The traction-free incremental boundary conditions~\eqref{EqBC-dotS} can be re-expressed through $\tensOldroydK$ as
\begin{equation}\label{EqBC-K-OldroydK}
    \OldroydK_{ir} = 0 \,\,\,\,\, \mbox{as}\,\,\,\,  \rred = \pm t/2 \quad( i = r, \theta, z)
\end{equation}
and a new set of generalized stresses is obtained from the initial definition, Eqs.~\eqref{Eq-Def-Generalized-FM}. 
In fact, by introducing the components of $\tensOldroydK$ given by Eqs.~\eqref{Eq-Def-tensOldroydK}, the first three Eqs.~\eqref{Eq-Integral-Incremental-Equilibrium-Final} are given the following \lq spatial' format:
\begin{equation}\label{Eq-Integral-Incremental-Equilibrium-Cylinder-Oldroyd-02}
     \begin{dcases}
           \begin{aligned}
          \,\pardd{ \Oldroydmtheta }{\theta}
            &  + a \,\pardpard{ ( \Oldroydmthetaz +\Oldroydmztheta)}{\theta}{z}
                + a^2 \, \pardd{ \Oldroydmz }{z}
                + a \, \Oldroydntheta \, +\\
            &\qquad\,\,   - P \, a^2 / t \intrred{ \left[ \pardpardd{\vtheta}{\theta}{z} \, \rred/a + \pardddz{ \vz } \; \rred + \parddz{\vr} \right] \, \left(1+\rred/a \right) }
               = 0  \comma
          \end{aligned} \\
          \, a \, \pardtheta{\Oldroydntheta}
          + a^2 \, \pardz{\Oldroydnthetaz}
          - \pardtheta{\Oldroydmtheta}
          - a \, \pardz{\Oldroydmthetaz}
          + P \, a^2 / t \intrred{ \parddz{\vtheta} \, {\left(1+\rred/a \right)}^2 }
          = 0 \comma\\
          \, a \, \pardz{\Oldroydnz}
          + \, \pardtheta{\Oldroydnztheta}
          + P \, a / t \intrred{ \parddz{\vz} \, \left(1+\rred/a \right) }
          = 0\point
     \end{dcases}
\end{equation}
%
%
\subsection{Rotational equilibrium about axis \texorpdfstring{$r$}{r}}
     \label{SubSec-Sixth-Equilibrium-Equation}
A sixth incremental equilibrium equation expressing the rotational equilibrium about axis $r$ can be obtained from 
a through-thickness integration of \EqRef{Eq-dotS-Component-Transformation}$_3$ after multiplication by $(1 + \rred/a)$
\begin{equation}
	\intrred{ \dotSthetaz \, \left(1+\rred/a \right) }\, - \intrred{ \dotSztheta \, \left(1+\rred/a \right) } - P/t \intrred{\pardz{\vtheta} \, \left(1+\rred/a \right) } =0\point 
\end{equation}
The introduction of the generalized stresses in~\EqRef{Eq-Def-Generalized-FM} leads to the expression in material formulation
\begin{equation}
\label{gnaccia}
a \, (\dotnthetaz - \dotnztheta) + \dotmztheta = P \, a/t \, \intrred{ \pardz{\vtheta} \, \left(1+\rred/a \right)} \comma
\end{equation}
while the spatial version in terms of Oldroyd increments reads
\begin{equation}
\label{Eq-Sixth-Equation-Like-Fluegge}
	a (\Oldroydnthetaz - \Oldroydnztheta) + \Oldroydmztheta = 0 \point 
\end{equation}

Note that all equations obtained until now, in particular Eqs.~\eqref{Eq-Integral-Incremental-Equilibrium-Final}, (\ref{Eq-Integral-Incremental-Equilibrium-Cylinder-Oldroyd-02}), (\ref{gnaccia}), and (\ref{Eq-Sixth-Equation-Like-Fluegge})  
do not involve any approximation and thus are rigorous.
%
%
\subsection{The \texorpdfstring{\Fluegge}{Fluegge} approximation}
As already mentioned, all equations derived so far, to be used in the following elaboration, \textit{are exact}. 
Interestingly, the corresponding equations provided by \textcite{Flugge1962} can be recovered as an approximation of Eqs.~\eqref{Eq-Integral-Incremental-Equilibrium-Cylinder-Oldroyd-02}, when the assumption is introduced that the cylinder wall thickness $t$ is small. In fact, a  Taylor series expansion allows to show that 
\begin{equation*}
     \begin{array}{l}
         \ds \frac{1}{t} \intrred{ v_{i} \, \rred/a  \, \left(1+\rred/a \right) } = \bigO{t^2/a^2}
     \end{array}
\end{equation*}
and therefore the equations introduced by  \textcite{Flugge1962} are recovered:
\begin{equation} \label{Eq-Integral-Incremental-Equilibrium-Cylinder-Oldroyd-02-Thin}%
     \begin{dcases}
          \pardd{ \Oldroydmtheta }{\theta}
               + a\pardpard{ ( \Oldroydmthetaz + \Oldroydmztheta)}{\theta}{z}
               + a^2 \, \pardd{ \Oldroydmz }{z}
               + a \, \Oldroydntheta
               - P \, a^2 / t \, \intrred{ \parddz{\vr} \, \left(1+\rred/a \right) }
            \approx 0 \comma \\
          a \, \pardtheta{\Oldroydntheta}
               + a^2 \, \pardz{\Oldroydnthetaz}
               - \pardtheta{\Oldroydmtheta} - a \, \pardz{\Oldroydmthetaz}
               + P \, a^2 / t \, \intrred{ \parddz{\vtheta} \, \left(1+\rred/a \right) }
         \approx  0 \comma \\
          a \, \pardz{\Oldroydnz}
          +  \, \pardtheta{\Oldroydnztheta}
          + P \, a / t \, \intrred{ \parddz{\vz} \, \left(1+\rred/a \right) }
          = 0\point
     \end{dcases}
\end{equation}
In addition to the above equations, \Fluegge used also Eq.~(\ref{Eq-Sixth-Equation-Like-Fluegge}), albeit he did never explicitly mention the use of either the \Oldroyd increment or the \Kirchhoff stress measure.
%
%
%
%
\section{Incremental deformations of a prestressed shell}
     \label{Sec-Kinematics}
As a premise, the Euler-Bernoulli  beam theory is briefly discussed on the basis of the standard assumptions \parencite{Love1906}.
The kinematics of a beam in a plane is described through the displacement $\bar{{\bf u}}(x_{01})$ of a generic point lying on its centroidal axis, singled out by the material coordinate $x_{01}$ along the straight reference configuration, $\overbar{{\bf x}}_0=x_{01}\e{1}$. Assuming that the centroidal axis behaves as the Euler's elastica corresponding to the evolution of an extensible line, the unit normal vector $\vecnmid$ (counterclockwise rotated $\pi/2$ with respect to the tangent) at point $\overbar{{\bf x}}=\left[x_{01}+\overbar{u}_{1}(x_{01})\right]\e{1}+ \overbar{u}_{2}(x_{01})\,\e{2}$ reads (\textcites{Bigoni2012}{Bigoni2019})
\begin{equation}
\label{Eq-Def-Unit-Normal-Deformed-Beam}
     \vecnmid(x_{01}) = \frac{-\umidp{2} \, \e{1}+ (1 + \umidp{1}) \, \e{2} }{\sqrt{(1 + \umidp{1})^2 + {\umidp{2}}^2}}\point
 \end{equation}

For any point of the beam in its spatial configuration, ${\bf x}=x_{1}\e{1}+ x_{2}\,\e{2}$, having ${\bf x}_0=x_{01}\e{1}+ x_{02}\,\e{2}$ as material counterpart  with $x_{02} \in [-t/2,+t/2]$, the following displacement field is postulated:
\begin{equation} \label{Eq-Def-Kinematics-Beam}
     \vecu(x_{01}, x_{02}) = \vecumid (x_{01}) + [\vecnmid (x_{01}) - \e{2}]\, x_{02} \point
\end{equation}
If the derivatives of the displacement components \eqref{Eq-Def-Kinematics-Beam} are negligible compared to  unity (i.e. ${\umidp{1}}$ and $\umidp{2} \ll\! 1$), the linearized kinematics of the Euler-Bernoulli  beam is recovered, namely,
\begin{equation}
          u_1(x_{01}, x_{02}) \approx \umid_1(x_{01}) - \umidp{2}(x_{01}) \,  x_{02}, \quad u_2(x_{01}, x_{02}) \approx \umid_2(x_{01})\point 
     \end{equation}
The kinematics of the incremental deformations in a prestressed cylindrical shell is illustrated as an extension of the development outlined above for the beam, following the standard assumptions discussed, among others, by  \textcites{Love1906}{Flugge1932}{Podio-Guidugli1989}{Steigmann2014}{Geymonat2007}.
In a cylindrical coordinate system, the prestressed shell configuration and its evolution after superposition of an incremental deformation are described through the geometry of the midsurface
(\textcites{Malvern1969, Ogden1984, Chapelle2011}), respectively as 
\begin{equation}
          \overbar{\vec{x}} = a \, \er + z \, \ez,  \quad
          {\overbar{\vec{x}}} ' = (a + \vmidr) \, \er + \vmidtheta \, \etheta + (z + \vmidz) \, \ez ,
     \end{equation}
where $a$ is the radius of the prestressed cylindrical midsurface, while $\vmidr(\theta,z)$, $\vmidtheta(\theta,z)$ and $\vmidz(\theta,z)$ represent its incremental displacement components.
The unit normal to the deformed surface is defined as
\begin{equation}\label{Eq-Def-Unit-Normal-Deformed-Shell}
     \vecnmid(\theta,z) = \frac{ \pardtheta{\overbar{\vec{x}} '\!} \times \pardz{\overbar{\vec{x}} '\!} }{ \abs{ \pardtheta{\overbar{\vec{x}} '\!} \times \pardz{\overbar{\vec{x}} '\!}} } \comma
\end{equation}
where $|\cdot|$ represents the norm of its vector argument, while the derivatives read
as 
\begin{equation}
\label{Eq-Unit-Tangent-Vectors}
          \pardtheta{\overbar{\vec{x}} '\!}
           =(\pardtheta{\vmidr} - \vmidtheta) \, \er + (a + \vmidr + \pardtheta{\vmidtheta}) \, \etheta + \pardtheta{\vmidz}\, \ez , \quad
          \pardz{\overbar{\vec{x}} '\!}= \pardz{\vmidr} \, \er + \pardz{\vmidtheta} \, \etheta + (1 + \pardz{\vmidz}) \, \ez \point 
\end{equation}
It is useful to consider two new vectors parallel to $\pardtheta{\vec{x}'}$ and $\pardz{\vec{x}'}$, respectively,
\begin{equation}\label{Eq-Unit-Tangent-Vectors-Normalized}
          \pardtheta{\hat{\vec{x}} '\!} = \ds
               \frac{( \pardtheta{\vmidr} - \vmidtheta) / a}{ 1 + (\vmidr + \pardtheta{\vmidtheta})/a } \, \er + \etheta + \frac{\pardtheta{\vmidz}/a}{1 + (\vmidr + \pardtheta{\vmidtheta})/a } \, \ez ,
               \quad
               \ds
               \pardz{\hat{\vec{x}} '\!} = \frac{\pardz{\vmidr}}{1 + \pardz{\vmidz}} \, \er + \frac{\pardz{\vmidtheta}}{1 + \pardz{\vmidz}} \, \etheta + \ez .
\end{equation}
Up to the leading-order, assuming the incremental displacement components $\vmidr$ and $\vmidtheta$ to be small (negligible if compared to radius $a$), and the incremental displacement gradient to be negligible with respect to unity, the following approximations can be introduced
\begin{equation}\label{Eq-Unit-Tangent-Vectors-Approximated}
         \pardtheta{\hat{\vec{x}} '\!}\approx
               \ds
               \frac{ \pardtheta{\vmidr} - \vmidtheta }{a}\, \er + \etheta + \frac{\pardtheta{\vmidz}}{a } \, \ez ,%
               \quad
               \pardz{\hat{\vec{x}} '\!}\approx \pardz{\vmidr}\, \er + \pardz{\vmidtheta}\, \etheta + \ez ,
\end{equation}
while the unit normal to the cylindrical surface $\vecnmid$ follows as
\begin{equation} \label{Eq-Unit-Normal-Result}
     \vecnmid \approx  \er + \frac{\vmidtheta - \pardtheta{\vmidr}}{a} \, \etheta - \pardz{\vmidr} \, \ez.
\end{equation}

Paralleling  the beam theory assumption,  Eqn. \eqref{Eq-Def-Kinematics-Beam}, the incremental kinematics of a cylindrical shell is represented in the form  \parencite{Chapelle2011}
\begin{equation} \label{Eq-Kinematics-Shell}
          \vecv(\rred,\theta,z) = \vecvmid(\theta,z) + \left[\vecnmid(\theta,z) - \e{r}\right] \rred\point 
\end{equation}
On the basis of the above-described  linearized kinematics, the gradient of the incremental displacement becomes 
\begin{equation}
	\label{Eq-Gradv}
     \tensL = \begin{bmatrix}
     	0                                           & (\pardtheta{\vmidr} - \vmidtheta)/a                                                                                                  & \pardz{\vmidr}                                                                                 \\[3mm]
     	(\vmidtheta - \pardtheta{\vmidr})/a & \ds  \left[ \vmidr - \rred/a \, \parddtheta{\vmidr} + \left( 1 + \rred/a \right)  \pardtheta{\vmidtheta} \right]/(a+\rred) & \ds \left( 1+ \rred / a \right)\pardz{\vmidtheta} - \rred / a\, \pardpard{\vmidr}{\theta}{z} \\[3mm]
     	\ds%
     	-\pardz{\vmidr}                  & \ds \left[ \pardtheta{\vmidz}  -  \rred \, \pardpard{\vmidr}{\theta}{z} \right]/(a+\rred)                                 & \pardz{\vmidz} - \rred \, \parddz{\vmidr}
     \end{bmatrix} ,
\end{equation}
so that the components of the Eulerian strain increment tensor \mbox{$\tensD = (\tensL+\tensL^\T)/2$} are 
\begin{equation}\label{Eq-Eulerian-Strain-Rate-Kinematics}
     \begin{array}{l}
           \ds 
           \Dr = 0 \comma\quad 
           \Dtheta = \left[ \vmidr - \rred/a \, \parddtheta{\vmidr} + ( 1 + \rred/a ) \, \pardtheta{\vmidtheta} \right]/(a + \rred) \comma \quad 
          \Dz =  \pardz{\vmidz}- \rred \, \parddz{\vmidr}\comma
          \\[3mm]
          \ds 
          \Drtheta =\Drz = 0 \comma\quad
          \Dthetaz = \left[- \rred/a\, (2 + \rred/a ) \, \pardpard{\vmidr}{\theta}{z} +{( 1 +\rred/a )}^2 \, \pardz{\vmidtheta}  +
          \pardtheta{\vmidz}/a\right] /\left(2\,(1+\rred/a)\right)\point
     \end{array}
\end{equation}
%
%
\section{Two constitutive equations for compressible hyperelasticity}
\label{Sec-Constitutive-Equations}
Two hyperelastic material models, both isotropic in their undeformed state, are considered, for which the strain energy functions are provided by \textcite[their Eqs.~(2.11) and~(2.12)]{Pence2015}. Adopting the same notation proposed by those authors, the strain energy functions $W_a$ and $W_b$ are adopted, namely,
\begin{equation}
     \label{Eq-Def-Energy-PenceGou2015}
     W_a = \frac{\mu}{2} \bigl[ \Invariant{1,\tensB} - 3 - \ln{\Invariant{3,\tensB}} \bigr] + \left(\frac{\kappa}{2} - \frac{\mu}{3}\right) {\left( \sqrt{\Invariant{3,\tensB}} - 1 \right)}^2 \comma
\end{equation}
and
\begin{equation}
     \label{Eq-Def-Energy-PenceGou2015_separable}
     W_b = \frac{\mu}{2} \left(\frac{\Invariant{1,\tensB}}{\Invariant{3,\tensB}^{1/3}} - 3 \right) + \frac{\kappa}{8} \left(\Invariant{3,\tensB}+ \frac{1}{\Invariant{3,\tensB}}-2\right),
\end{equation}
where $\Invariant{1,\tensB} = \tr{\tensB}$ and $\Invariant{3,\tensB} = \det{\tensB}$, while $\mu$ and $\kappa$ represent the shear and bulk moduli of the material in its unstressed state, related to the \Young modulus $E$ and \Poisson's ratio $\nu$ through the usual formulae, namely,  \mbox{$\mu = E/(2 \, (1+\nu))$} and \mbox{$\kappa = E/(3 \, (1-2\nu))$}.

The strain energy function~\eqref{Eq-Def-Energy-PenceGou2015} is a special form of the general Blatz-Ko material model, in contrast with the strain energy function (\ref{Eq-Def-Energy-PenceGou2015_separable}), which allows instead a separation between the pure volumetric effects and other contributions from the deformation. 
Both the models describe compressible neo-Hookean materials and satisfy, in the undeformed state, the stress-free condition,
as well as the consistency with the classical linearized elasticity theory. 
Therefore, with reference to the generic strain energy function $W$, the following conditions hold true
\begin{equation} 
\label{Eq-Energy-Consistency-Conditions-Horgan} \pcomma
     \begin{dcases}
          \pard{\overbar{W}}{1} + 2 \pard{\overbar{W}}{2} + \pard{\overbar{W}}{3} = 0 \comma \\
          \pard{\overbar{W}}{1} + \pard{\overbar{W}}{2} = -(\pard{\overbar{W}}{2} + \pard{\overbar{W}}{3}) = \mu / 2 \comma \\
          \pardd{\overbar{W}}{1} + 4 \pardpard{\overbar{W}}{1}{2} + 4 \pardd{\overbar{W}}{2} + 2 \pardpard{\overbar{W}}{1}{3} +  4 \pardpard{\overbar{W}}{2}{3} + \pardd{\overbar{W}}{3} = \kappa / 4 + \mu / 3 ,
     \end{dcases}
\end{equation}
where the derivatives $\pard{\overbar{W}}{i} = \partial{}{W(\Invariant{1}, \Invariant{3})}/\partial{}{\Invariant{i}}$ are to be evaluated for $\Invariant{1} = \Invariant{2} =  3$ and $\Invariant{3} = 1$ \parencite{Horgan2004}. 

The \Cauchy stress, in general defined according to 
\begin{equation}
	\label{Def_T_from_W}
	\begin{array}{l}
		\tensT = 2  J^{-1} (W,_1\tensB+\Invariant{3}W,_3\tensI)\comma 
        \end{array}
\end{equation}
assumes for the strain energy (\ref{Eq-Def-Energy-PenceGou2015}) the expression
\begin{equation}
	\label{Eq-TensK-TensS-General}
	\begin{array}{l}
		\tensT_a = \mu J^{-1}\, (\tensB - \tensI) + (\kappa - 2/3\, \mu) (J - 1)\, \tensI  \comma
        \end{array}
\end{equation}
while, for the strain energy (\ref{Eq-Def-Energy-PenceGou2015_separable}), it reads as
\begin{equation}
	\label{Eq-TensK-TensS-General_b}
	\begin{array}{l}
		\tensT_b = \mu J^{-5/3}\, \left(\tensB - I_1/3\;\tensI\right) + \kappa/4 \left(J^4-1\right) J^{-3}\,\tensI .
        \end{array}
\end{equation}

Through the relative Lagrangean description, in which the current configuration is assumed as reference, the Oldroyd increment of the \Kirchhoff stress turns out to be related to the strain energy density of a hyperelastic material $W$ as \parencite{Bigoni2012} 
\begin{equation}
 \label{Eq-tensOldroydK-From-Energy} \pcomma
     \tensOldroydK  = \tenf{H}[\tensD] =J^{-1} \, ( \tensF \boxtimes \tensF )\,  \frac{\partial^2{W}}{\partial{\tensE}^2}\, ( \tensF \boxtimes \tensF )^T
     \left[\tensD\right]\!\comma  
\end{equation}
where $\tensE$ denotes the \GL strain tensor, while the tensor product $\boxtimes$ is defined as $(\tensA \boxtimes \tens{B}) [\tens{C}] = \tensA \tens{C} \tens{B}^{\T}$.
Inserting the form~\eqref{Eq-Def-Energy-PenceGou2015} for the strain energy function $W_a$ into \EqRef{Eq-tensOldroydK-From-Energy}, the following expression for the elastic fourth-order tensor $\tenf{H}$ is derived 
\begin{equation*}
     \tenf{H}_a = (\kappa - 2/3 \mu) (2 J - 1) \, \tensI \otimes \tensI 
          + 2 \left[ \mu/  J - (\kappa - 2/3 \mu) (J - 1) \right] \,  \tenfS,
\end{equation*}
where $\tenfS$ is the fourth-order symmetrizer tensor, leaving $\tensD$ unchanged because of symmetry, namely, $\tenfS[\tensD]=(\tensD+\tensD^T)/2=\tensD$.
Therefore the Oldroyd increment of the Kirchhoff stress~\eqref{Eq-TensK-TensS-General} for the model with strain energy (\ref{Eq-Def-Energy-PenceGou2015}) becomes 
\begin{equation}
	\label{Eq-tensOldroydK-From-Energy-PenceGou}
	\tensOldroydK_a = (\kappa - 2/3 \, \mu) (2 J - 1) (\tr{\tensD})\,\, \tensI + 2 \left[ \mu/J - (\kappa - 2/3\, \mu) (J - 1) \right] \tensD,
\end{equation}
while, paralleling the procedure for the model with strain energy $W_b$, Eq.~(\ref{Eq-Def-Energy-PenceGou2015_separable}), the following expression is obtained:  
\begin{equation}
	\label{Eq-tensOldroydK-From-Energy-PenceGou_b}
	\tensOldroydK_b =\left[\left(\frac{2\, \mu\, I_1}{9\, J^{5/3}}+\frac{\kappa}{2}\,\frac{J^4+1}{J^3}\right)\,\tensI-\frac{2\, \mu }{3\,J^{5/3}}\tensB\right]\tr{\tensD}  - \frac{2\, \mu }{3\, J^{5/3}}(\tr{\tensB\tensD})\,\tensI 
+\left(\frac{2\, \mu\, I_1}{3\, J^{5/3}}-\frac{\kappa}{2}\, \frac{J^4-1}{J^3}\right)\tensD \point
\end{equation}
It is noteworthy to point out that {\it the constitutive equation used by \Fluegge can be recovered from both Eqs. (\ref{Eq-tensOldroydK-From-Energy-PenceGou}) and (\ref{Eq-tensOldroydK-From-Energy-PenceGou_b})}, assuming the pre-stressed and unstressed configurations to be coincident, so that $\tensF = \tensI$, a condition leading to 
\begin{equation}\label{vecio-flugge}
     \tensOldroydK = (\kappa - 2/3 \, \mu) \left(\tr{ \tensD } \right)\, \tensI + 2 \mu \tensD, 
\end{equation}
which represents the incremental law used by \Fluegge.
%
\subsection{Axisymmetric pre-stress}
%
The axisymmetric ground-state assumption prescribes the coincidence of the radial and circumferential stretches, $\lambdar=\lambdatheta$, as well as the vanishing of the radial and circumferential stress components, $\Tr = \Ttheta= 0$. 
Therefore, with regard to strain energy function $W_a$, the following condition is obtained from \EqRef{Eq-TensK-TensS-General} for the 
components of the Cauchy stress in the trivial configuration:
\begin{equation}
\label{EqKr=0}\ppoint
      T_{a_{rr}} = T_{a_{\theta\theta}} =\mu \, (\lambdar^2 - 1)/(\lambdar^2 \, \lambdaz) + (\kappa - 2/3\, \mu) \,  (\lambdar^2 \, \lambdaz - 1)  = 0 \point
\end{equation}
Solving \EqRef{EqKr=0} for the radial stretch $\lambdar$ yields
\begin{equation}
\label{Eq-Lambdar-Sol}
          \lambda_r =  \sqrt{{\frac{2 \nu \, (\lambdaz+1) +\delta-1}{4\nu\,\lambdaz^2}}} \comma
\end{equation}
where $\delta = \sqrt{1+4\, \nu (\lambdaz-1) \left[(2-3 \nu) \lambdaz +1 - \nu  \right] }$. Note that both $\delta$ and $\lambda_r$ are real for $\nu \in [0, \,0.5]$.
The stress tensor in Eq.~\eqref{Eq-TensK-TensS-General} can be simplified by means of \EqRef{Eq-Lambdar-Sol}, so that its only nonzero component turns out to be  
\begin{equation}
 \label{Eq-Sz-Kz-Tz-Sol}
               T_{a_{zz}} = \mu \,\frac{2\, \nu\lambda _z \left(2\lambda _z^3-1\right) -\delta+1-2\, \nu}{\lambda _z (2\, \nu(\lambda _z+1)+\delta-1)}\point
\end{equation}

A substitution of \EqRef{Eq-Lambdar-Sol}  into \EqRef{Eq-tensOldroydK-From-Energy-PenceGou} yields the following expressions for the diagonal (denoted assuming the repeated indices $i$ not to be summed over) and the out-of-diagonal components of the Oldroyd increment of the Kirchhoff stress, tensor $\tensOldroydK$:
\begin{equation}
	\label{Eq-tensK-PS}
\qquad {\OldroydK}_{a_{ii}} =\frac{\mu \left[\left(1-2\nu)(2D_{ii}-\tr{ \tensD } \right)+\delta\tr{ \tensD }\right]}{(1-2\nu)\,\lambda_z}\comma
 \quad     
 {\OldroydK}_{a_{ij}} = \frac{2 \mu\, D_{ij}}{\lambdaz} 
 \qquad
 (i,j =  r, \theta, z , \,\, i \neq j) \point
\end{equation}

Similarly to Eq.~(\ref{EqKr=0}), the condition of axisymmetric pre-stress for a material admitting the strain energy function $W_b$, \EqRef{Eq-Def-Energy-PenceGou2015_separable}, can be written as
\begin{equation}
\label{EqKrb=0}
 T_{b_{rr}} = T_{b_{\theta\theta}} = \mu\,\left(\lambdar^2-\lambdaz^2\right)/\left(3\,\lambdar^{10/3} \lambdaz^{5/3}\right) +
 \kappa\,\left(\lambdar^8 \lambdaz^4-1\right)/\left(4\, \lambdar^6 \lambdaz^3\right) =0,
\end{equation}
which can be solved numerically to compute $\lambda_r$ as a function of the pre-stretch $ \lambda_z$ for $\nu \in [0, \,0.5)$. The corresponding axial stress $T_{b_{zz}}$, as well as the components of the Oldroyd increment of the Kirchhoff stress  ${\OldroydK}_{b_{ij}}$ are finally evaluated by means of \EqRef{Eq-TensK-TensS-General_b} and \EqRef{Eq-tensOldroydK-From-Energy-PenceGou_b}, respectively.

Noteworthy, in the incompressible limit, $\nu\to 0.5$, the radial stretch tends to the incompressibility constraint $\lambda_r=\lambda_z^{-1/2}$ with dimensionless axial stress $T_{zz}/\mu=(\lambda_z^3-1)/\lambda_z$ for both materials with the strain energy functions $W_a$ and $W_b$,  \EqRef{Eq-Def-Energy-PenceGou2015} and \EqRef{Eq-Def-Energy-PenceGou2015_separable}. 

\begin{figure}[t!]
	\centering
	\includegraphics[width=0.98\textwidth]{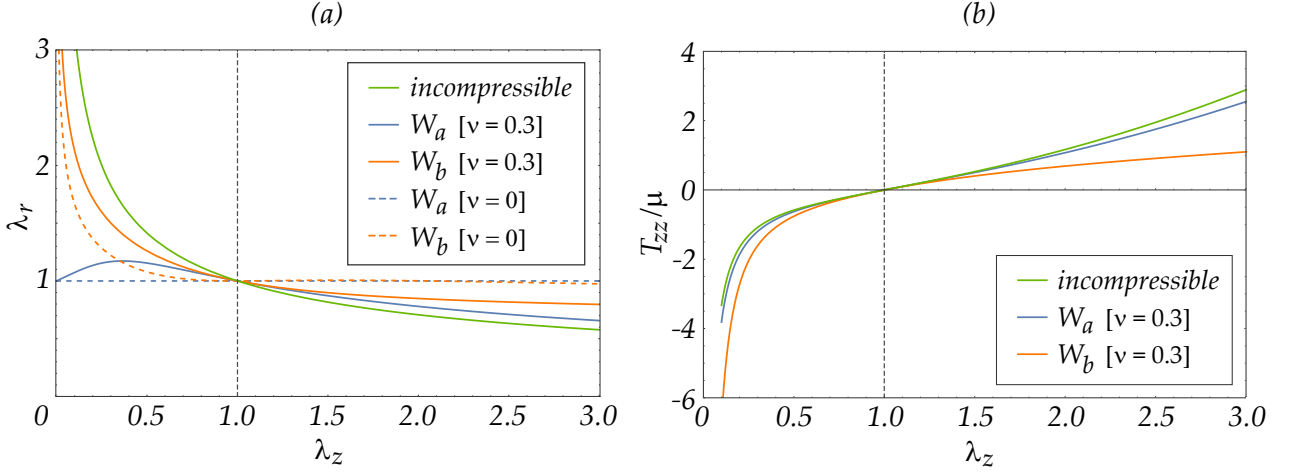}
	\vspace*{.2cm}
	\caption{Uniaxial loading before bifurcation for a compressed cylinder,  following from two the elastic models with the strain energies $W_a$  (blue lines) and $W_b$  (red lines). (a) The radial stretch $\lambda_r$ is determined as a function of the axial stretch  $\lambda_z$; 
	$\nu = \{0,0.3\}$ are considered, together with the incompressibility limit, $\nu = 0.5$.
	(b) The axial Cauchy stress $T_{zz}/\mu$ is determined as a function of the axial stretch $\lambda_z$; $\nu=0.3$ and $\nu = 0.5$ are considered. 
	}
	\label{fig:constitutive_laws}
\end{figure}

The radial stretch $\lambda_r$ and the dimensionless axial stress $T_{zz}/\mu$ 
are reported in 
\FigureRef{fig:constitutive_laws}, for the two models characterized by the strain energies $W_a$ (blue lines) and $W_b$ (red lines), as functions of the axial stretch $\lambda_z$. 
For the strain energy function $W_a$, 
Eqs. (\ref{Eq-Lambdar-Sol}) 
and  (\ref{Eq-Sz-Kz-Tz-Sol}) have been used, while for the strain energy function $W_b$, 
Eq.~(\ref{EqKrb=0}) has to be numerically solved. 
Three values of $\nu$ are reported in \FigureRef{fig:constitutive_laws} (a), namely, $\nu=0.3$ (continuous lines), $\nu = 0$ (dashed lines) and the limit $\nu=0.5$ (green line) corresponding to incompressibility, where the two models provide the same response. Only two values of $\nu$, namely 0.3 and 0.5 are reported in \FigureRef{fig:constitutive_laws} (b).

The curves demonstrate the high non-linearity of the models and the differences in the mechanical response to stretch. Note that when $\nu = 0$, the radial stretch is constant and equal to unity, $\lambda_r=1$, for the strain energy $W_a$, while for $W_b$, $\lambda_r$ remains close to 1 for values of $\lambda_z>0.7$. 
%
\subsection{Incremental plane stress assumption}
%
For cylinders having `sufficiently' thin walls, the assumption of plane stress becomes reasonable and is hereafter extended to the bifurcation state as well,  namely,
\begin{equation}\label{Eq-dotSr=0}\ppoint
     \dot{S}_{rr} = 0 \comma \quad \forall\rred \in [-t/2; \,t/2] \point
\end{equation}
Recognizing that $\OldroydK_{rr}=\dot{S}_{rr}$ as a result of the assumed structure of the pre-stress in \EqRef{Eq-Def-tensK}, together with \EqRef{Eq-Def-tensOldroydK}, the enforcement of \EqRef{Eq-dotSr=0} for the material with the strain energy function $W_a$ in \EqRef{Eq-Def-Energy-PenceGou2015}, yields\\
\begin{equation}
\label{Eq-DrWa}
    \Dr=\, \frac{ \nu\, \lambdar^2\, \lambdaz \left(2 \lambdar^2\, \lambdaz-1\right)}
    {\nu \left(2-\lambdar^2\, \lambdaz\right)-1}\left(\!\Dtheta+\Dz\right)\!\comma
\end{equation}
to be further simplified through the introduction of \EqRef{Eq-Lambdar-Sol} as
\begin{equation}
\label{Eq-Dr}\ppoint
    \Dr=\, \frac{1 - 2\, \nu -\delta }{1 - 2 \,\nu + \delta }\, \left(\!\Dtheta+\Dz\right).
\end{equation}
Under the constraint represented by \EqRef{Eq-Dr}, the incremental constitutive equations \eqref{Eq-tensK-PS} assume the following expression:
\begin{equation}
\label{Eq-tensK-PS-Non-Zero-Components}
     \begin{array}{c}
          \ds \OldroydK_{a_{rr}}=0\comma\quad
          \OldroydK_{a_{\theta\theta}} = \frac{2\mu\,\left[2\delta\,\Dtheta+(\delta-1+2\nu)\, \Dz\right]}{(\delta+1-2\nu)\,\lambda_z} \comma \quad
          \OldroydK_{a_{zz}}  = \frac{2\mu\,\left[(\delta-1+2\nu)\,\Dtheta+2\delta\,\Dz \right]}{(\delta+1-2\nu)\,\lambda_z}\comma\\[6mm]
     \OldroydK_{a_{r \theta}} = 2\mu\,\Drtheta / \lambdaz \comma \quad %
          \OldroydK_{a_{rz}} =  2\mu\,\Drz / \lambdaz \comma \quad%
          \OldroydK_{a_{\theta z}} =  2\mu\,\Dthetaz / \lambdaz \point
     \end{array}
\end{equation}

For the material with strain energy function $W_b$ in \EqRef{Eq-Def-Energy-PenceGou2015_separable}, the fulfillment of the plane stress requirement,  \EqRef{Eq-dotSr=0}, instead of \EqRef{Eq-Dr}, leads to %
\begin{equation}
\label{Eq-DrWb}\ppoint
\Dr=\,\frac{d_{\theta \theta} D_{\theta \theta}+d_{zz} D_{zz}}
{2 \left(2(1-2\nu ) \left(\lambda_r^2+2\lambda_z^2\right) \left(\lambda_r^2 \lambda_z\right)^{4/3}+3 (1+\nu)\right)},
\end{equation}
 where
\begin{equation*}
	\begin{gathered}
		d_{\theta \theta} = 2 (1-2 \nu) \left(4\lambda_r^2-\lambda_z^2\right) \left(\lambda_r^2\lambda_z\right)^{4/3}-
3 (1+\nu) \left(\lambda_r^8 \lambda_z^4+1\right),\\%
		d_{zz}=2 (1-2 \nu) \left(\lambda_r^2+ 2\lambda_z^2\right) \left(\lambda_r^2 \lambda_z\right)^{4/3}-
3 (1+\nu) \left(\lambda_r^8 \lambda_z^4+1\right)]\, .%
	\end{gathered}
\end{equation*}
Finally, the substitution of \EqRef{Eq-DrWb} into \EqRef{Eq-tensOldroydK-From-Energy-PenceGou_b}, after introducing the implicit relation $\lambda_r(\lambda_z)$ represented in \FigureRef{fig:constitutive_laws} (a) that aims to satisfy \EqRef{EqKrb=0}, allows to determine the components of tensor $\tensOldroydK_b$, whose expression remains in implicit form for the model with strain energy $W_b$.
%
%
\section{Bifurcation of an axially-compressed thin-walled cylinder}
\label{bif-cyl}
The bifurcation problem for an axially-compressed thin-walled cylinder is set up on the basis of the kinematical conditions~\eqref{Eq-Eulerian-Strain-Rate-Kinematics}, the
equilibrium equations~\eqref{Eq-Integral-Incremental-Equilibrium-Cylinder-Oldroyd-02}, expressed in terms of generalized incremental stresses, and the constitutive relations 
\begin{itemize}
\item{\EqRef{Eq-tensK-PS-Non-Zero-Components}, for the material obeying the strain energy function $W_a$},
\item{\EqRef{Eq-tensOldroydK-From-Energy-PenceGou_b} together with \EqRef{Eq-DrWb}, and the implicit relation $\lambda_r(\lambda_z)$ satisfying \EqRef{EqKrb=0}, for the material obeying the strain energy function $W_b$}.
\end{itemize}
The pre-stress load per unit length $P$ in equations~\eqref{Eq-Integral-Incremental-Equilibrium-Cylinder-Oldroyd-02} can be evaluated for the two materials by means of~\EqRef{Eq-TensK-TensS-General} and~\EqRef{Eq-TensK-TensS-General_b}, respectively.  

In the following of this article, explicit calculations will be presented with reference to the constitutive law following from the strain energy function $W_a$, \EqRef{Eq-Def-Energy-PenceGou2015}, with the index $a$ omitted (for the sake of conciseness). The analogous calculations we have performed for the function $W_b$ in \EqRef{Eq-Def-Energy-PenceGou2015_separable} are not reported here. Final computations of the bifurcation solution and asymptotic derivations of limit loads will be presented for both models.

As standard in the incremental bifurcation analysis of elastic solids (\cite{Hill1975}), the following ansatz is introduced for the incremental displacements at bifurcation, corresponding to a free sliding condition along perfectly smooth rigid constraints on the upper ($z=l$) and lower ($z=0$) faces: 
\begin{equation}\label{Eq-Ansatz-Velocity-Bifurcation}
     \begin{dcases}
          \vmidr (\theta,z) = c_1 \cos{(n \, \theta)} \cos{\left(\eta \, z/a\right)} \comma\\
          \vmidtheta (\theta,z) = c_2 \sin{(n \, \theta)} \cos{\left(\eta \, z/a\right)} \comma\\
          \vmidz (\theta,z) = c_3 \cos{(n \, \theta)} \sin{\left(\eta \, z/a\right)} ,
     \end{dcases}
\end{equation}
where $n=0,1,2,...$ and $\eta = m \pi a / l$  $(m=1,2,...)$ represent the circumferential and the longitudinal wave-numbers, respectively, singling out the bifurcation mode, while the amplitudes are collected in the vector $\vec{c} = \{c_1, c_2, c_3\}^T$.
The incremental displacement field,~\EqRef{Eq-Ansatz-Velocity-Bifurcation}, constant throughout the thickness of the shell, enforces the conditions of null incremental force $\Oldroydnthetaz$ and moment $\Oldroydmthetaz$ at the ends $z=0$ and $z=l$. In \textcites{Flugge1973}  the boundary conditions for the lower and upper ends were modelled as simple supports, therefore preventing radial and circumferential incremental displacements, while no restrictions were imposed on the axial incremental displacement. However, both the boundary conditions assumed by us and by \Fluegge lead to the same bifurcation conditions. 

Through the introduction of Eqs.~\eqref{Eq-Ansatz-Velocity-Bifurcation} into the kinematical conditions~\eqref{Eq-Eulerian-Strain-Rate-Kinematics} and the substitution into the constitutive relations~\eqref{Eq-tensK-PS-Non-Zero-Components},
the final form of the three incremental equilibrium equations~\eqref{Eq-Integral-Incremental-Equilibrium-Cylinder-Oldroyd-02} is obtained, with the generalized stresses defined according to Eqs.~\eqref{Eq-Def-Generalized-FM}. 
The bifurcation condition is eventually expressed in the standard form as
$\tens{M}\,\vec{c} = \vec{0}$,
where matrix $\tens{M}$ 
is a function of the axial stretch $\lambdaz$ (while $\lambda_r$ is replaced through \EqRef{Eq-Lambdar-Sol}), the dimensionless thickness of the shell $\tau = t/a$, the material parameter $\nu$ and the wave-numbers $n$ and $\eta$.
Bifurcation occurs when the coefficient matrix is singular, 
\begin{equation}\label{Eq-Determinant}
     \det{\tens{M}} = 0,
\end{equation}
a condition that allows to define the critical stretch $\lambdaz$ for bifurcation (and therefore the corresponding dimensionless axial compressive load $\pz = -P/D$, with \mbox{$D = E t / (1-\nu^2)$} representing the extensional stiffness of the shell), as a function of the geometrical variable $\tau$, the material parameter $\nu$ and the wave-numbers $n$ and $\eta$.
\begin{figure}[!htb]
	\centering%
	\includegraphics[width=0.85\textwidth]{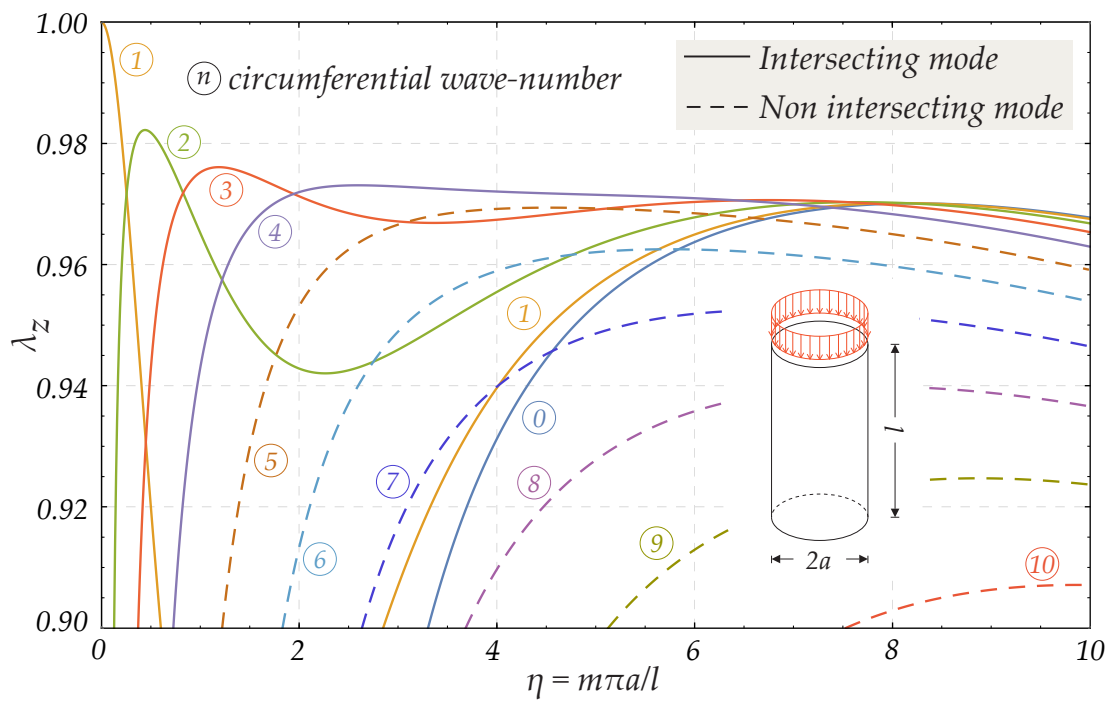}
    \vspace*{.1cm}
	\caption{Critical stretch $\lambdaz$ of an axially-compressed thin-walled cylinder ($\reoverri = 1.05$) made up of a Pence-Gou compressible material with strain energy function $W_a$ ($\nu = 0.3$) as a function of the longitudinal wave-number $\eta$: the curves for different values of the circumferential wave-number $n$ are denoted by {\Large \raisebox{-1.3 pt}{\textcircled{\raisebox{0.2 pt}{\small {$n$}}}}}. Continuous lines represent the intersecting critical modes contributing to the buckling envelope, while the dashed lines correspond to modes arising at  higher loads. The anti-symmetric mode labeled {\Large \raisebox{-1.3 pt}{\textcircled{\raisebox{0.2 pt}{\small {1}}}}} represents Euler buckling ($n=1$).}
\label{Plot-Buckling-Example-Envelope}
\end{figure}

\FigureRef{Plot-Buckling-Example-Envelope} shows the buckling diagram obtained for $\nu=0.3$ and $\reoverri = 1.05$, so that $\tau=0.0488$  
(note that both the radii ratio $\reoverri$ and the dimensionless thickness $\tau$ remain constant during the pre-bifurcation deformation, while 
the cylinder deforms maintaining its shape). The critical axial stretch is plotted as a function of the longitudinal wave-number $\eta$ for different values of the circumferential wave-number $n$. The critical modes are illustrated as continuous lines, while the dashed lines represent the modes corresponding to high axial stresses that cannot be reached when the load is continuously increased from zero. 
As expected, for small values of $\eta$, corresponding to very slender cylinders, the mode representative of the Euler buckling, characterized by $n=1$, becomes dominant. 

\begin{figure}[t!]
	\centering
	\includegraphics{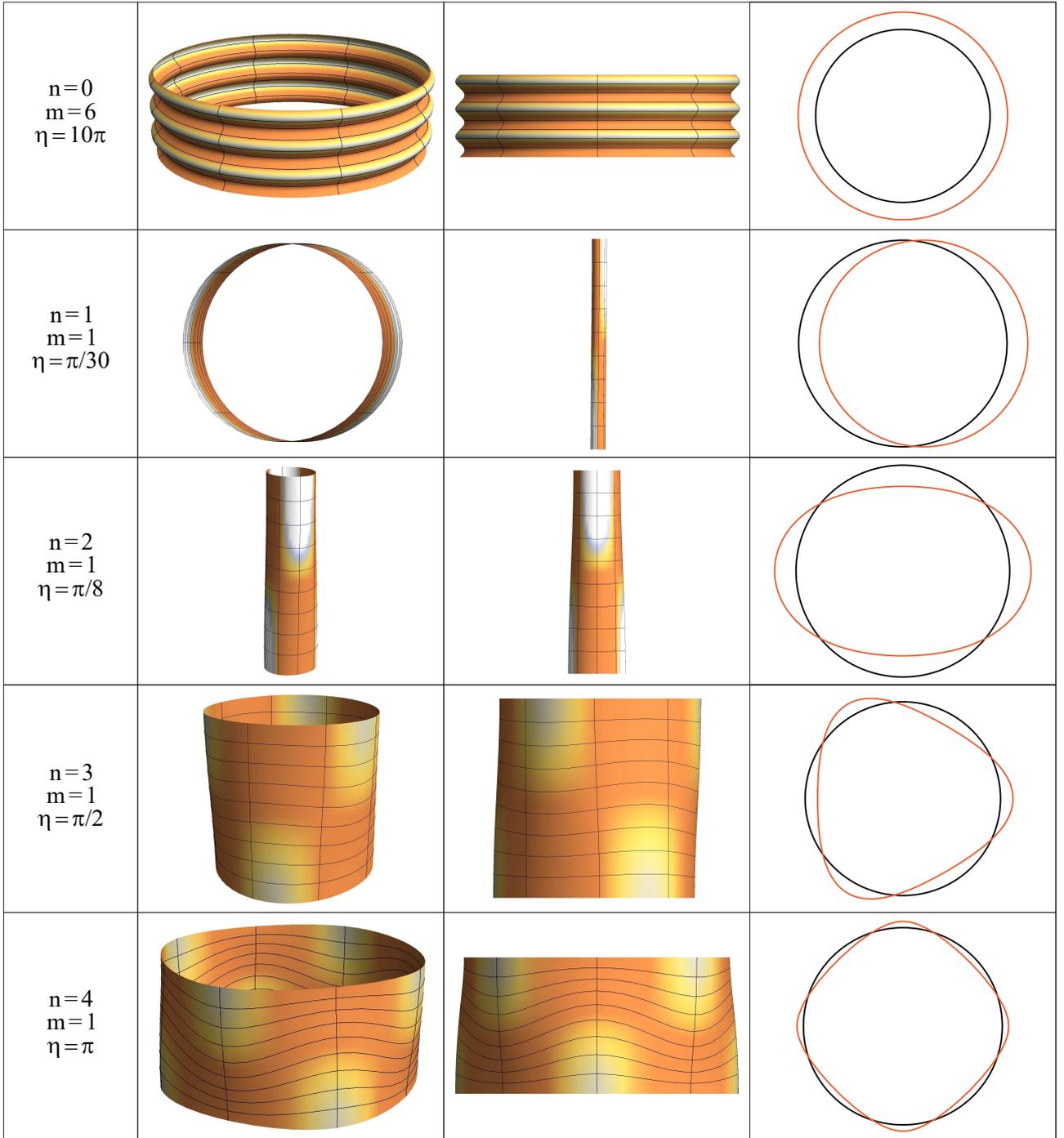}
	\caption{Different views for a selection of bifurcation eigenmodes for a shell with $\reoverri = 1.05$. The material obeys the Pence-Gou model with strain energy $W_a$ and $\nu = 0.3$. In particular, proceeding top-down, a surface instability mode, the Euler's buckling mode, and three different ovalization modes are shown.}
	\label{fig:eigenmodes}
\end{figure}

A selection of the bifurcation eigenmodes for a thin shell, corresponding to different values of circumferential and longitudinal wave-numbers $n$ and $m$, is displayed in \FigureRef{fig:eigenmodes}, where the colours highlight the peculiar bulges of the buckled shell geometry.

\begin{figure}[!htb]
	\centering
	\includegraphics[width=0.85\textwidth]{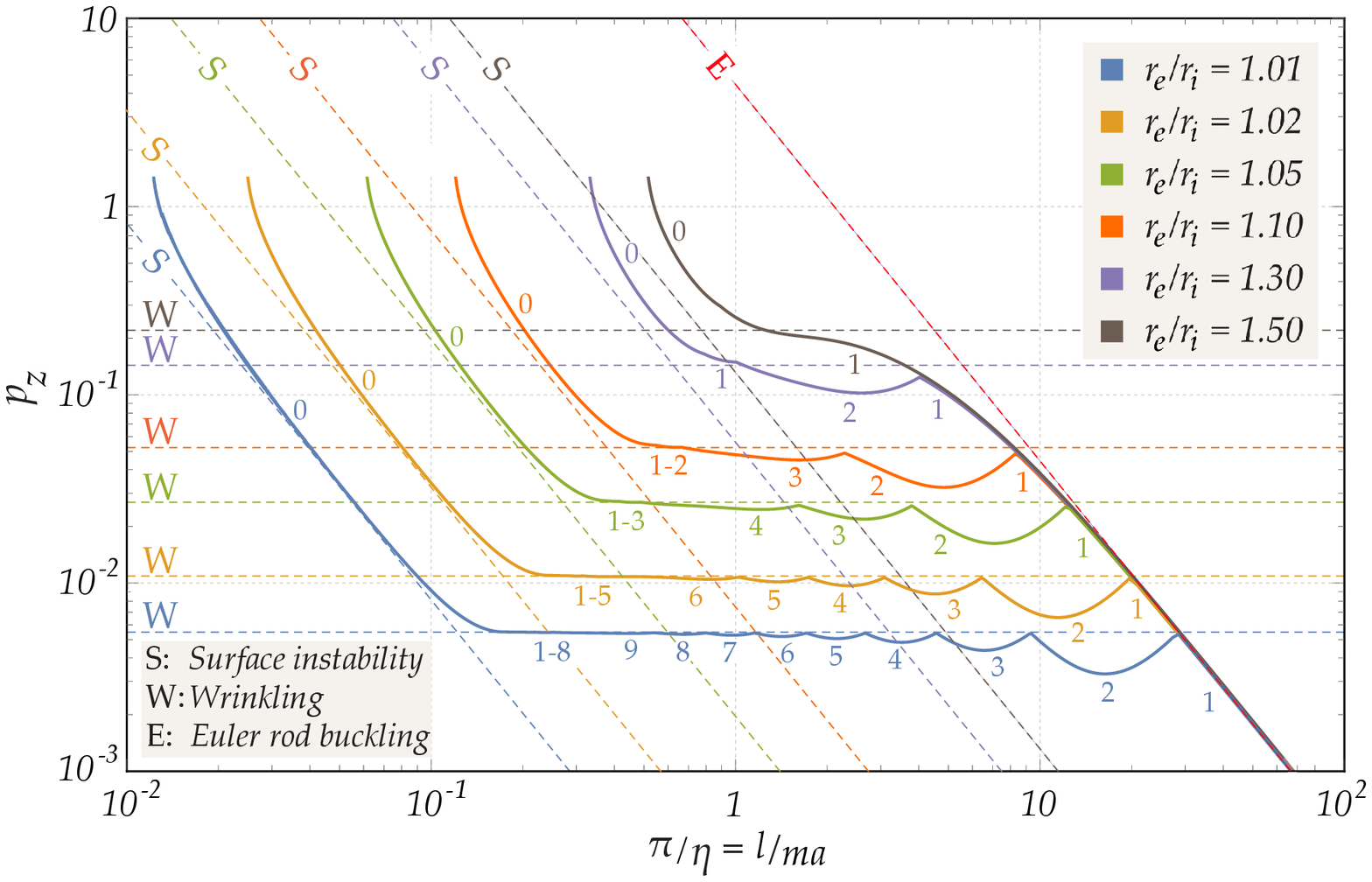}
	\vspace*{.1cm}
	\caption
	{Lower envelopes of the dimensionless load $\pz$ at bifurcation evaluated for a thin-walled cylinder made up of a Pence-Gou compressible material with strain energy $W_a$ ($\nu = 0.3$) as a function of $\pi/\eta=l/(ma)$ for different ratios $\reoverri$ between the external and internal radii of the cylinder (bi-logarithmic representation). The numbers adjacent to the curves indicate the critical circumferential modes of wave-number $n$, alternating along the envelopes for different values of $\pi/\eta$. The dashed lines illustrate the asymptotic buckling loads for surface instability (S), wrinkling (W) and the Euler's column (E).} 
\label{Plot-Buckling}
\end{figure}

Critical envelopes of the intersecting buckling curves are shown in \FigureRef{Plot-Buckling}, for different ratios $\reoverri$ and various circumferential wave numbers $n$. Depending on the ratio $l/(ma)$, the  bi-logarithmic plot shown in  \FigureRef{Plot-Buckling} highlights the sequence of three different behaviour ranges, found by \Fluegge and now recovered for two exact models of compressible elasticity:  
\begin{itemize} 
\item Cylinders with very small curvature (region on the left), tending to behave as plates, therefore the bifurcation condition approaches the plate buckling. 
\FigureRef{Plot-Buckling} highlights how the bifurcation solution pertaining to a thin plate (denoted by the letter $S$ in the figure), tends to progressively dissociate from the bifurcation solution for a thin-walled cylinder at increasing cylinder wall thicknesses. This analysis will be addressed in  Sect.~\ref{SubSec-plate};
\item  Moderately long cylinders (intermediate region) present an almost constant buckling load, independent of both the circumferential and longitudinal wave-numbers. This load is denoted in \FigureRef{Plot-Buckling} by the letter $W$ and analyzed in Sect.~\ref{SubSec-Mid-Long-Shell};
\item Cylinders with high slenderness (on the right) approach the Euler buckling solution, denoted in \FigureRef{Plot-Buckling} by the letter $E$. A detailed investigation of this case is presented in Sect.~\ref{SubSec-Euler-Buckling}.
\end{itemize}

The results presented above are based on a large strain approach with a constitutive equation assuming the strain energy $W_a$,~\EqRef{Eq-Def-Energy-PenceGou2015}.
We have obtained similar results with the strain energy $W_b$,~\EqRef{Eq-Def-Energy-PenceGou2015_separable}, not reported here for conciseness. Both cases are different from the small strain analysis performed by \Fluegge, which is based on a constitutive equation not following from a potential. %
Nevertheless, results in terms of critical loads for bifurcation turn out to be only marginally dependent on the constitutive equations, because bifurcation occurs at low stretch. Therefore, a comparison between the approach pursued in this paper and the solution obtained by \Fluegge shows almost coincident results; the comparison is not reported here as the curves are scarcely distinguishable. 

The accuracy of the current 2D approach (developed on the basis of two models provided by Pence-Gou and presented in \SectionRef{Sec-Constitutive-Equations}) will definitely be assessed though a comparison with the 3D full-field solution for bifurcation
on the basis of the constitutive model with strain energy $W_a$,~\EqRef{Eq-Def-Energy-PenceGou2015}
(\FigureRef{fig:cfr} in Sect.~\ref{Sec-3D-Solution}).
%
%
\section{Limiting cases via asymptotic analysis}\label{limiti}

Three crucial limiting cases are analyzed in this Section. The well-known solutions for cylinders with a very small curvature and for moderately long cylinders are rigorously derived from the finite elasticity approach developed in this article on the basis of both the constitutive models,  Eqs~\eqref{Eq-Def-Energy-PenceGou2015} and ~\eqref{Eq-Def-Energy-PenceGou2015_separable}. 
The problem of an Euler rod consisting in a  hollow cylindrical shaft is finally addressed. 

In all cases the asymptotic solutions are obtained for both the elasticity models considered here. Again, for the sake of conciseness, all the results will be presented only for the material whose strain energy function is $W_a$,~\EqRef{Eq-Def-Energy-PenceGou2015}.

At this point it is convenient to introduce the relationships expressing the push-forward operation
\begin{equation}
\label{vars_update}
     \tau = \frac{t}{a} = \frac{t_0}{a_0} = \tau_0\comma%
     \quad%
     \eta = m\pi \frac{a}{l} = m \pi  \, \frac{a_0\,\lambdar}{l_0\,\lambdaz} =\eta _0\,\frac{\lambdar}{\lambdaz}\comma%
\end{equation}
%
%
\subsection{Cylinders of very small curvature: surface instability}
\label{SubSec-plate}%

If the reference geometry of the shell is altered, increasing the radius $a_0$, while keeping constant both the length $l_0$ and the thickness $t_0$, a hollow cylinder of very small curvature is generated. The latter exhibits the surface instability of a plane plate strip with two free and two constrained opposite edges (endowed with 
simple supports, or, equivalently,  sliding clamps), subject to an in-plane dead load. The bifurcation solution for such plate strip is known, see  \textcites{Timoshenko1961Student}{Flugge1973}, 
\begin{equation} 
\label{Eq-Plate-Critical-Load}
	p_{z, \mathrm{S}} = k_0 \,\eta_0^2
\end{equation}
where $k_0=K_0/(D_0\,a_0^2)=\tau_0^2/12$, being $K_0=E \, t_0^3 / [12 \, (1 - \nu^2)]$ and $D_0=Et_0/ (1 - \nu^2)$ the flexural and extensional stiffnesses of the shell in its reference configuration (note that 
in current configuration,
$k=K/(D\,a^2)=k_0$). 
From the analysis of the critical pairs $\{\lambda_z,\eta\}$ obtained for the constitutive law  \EqRef{Eq-Def-Energy-PenceGou2015}, it turns out that, as recognized by \Fluegge, at high values of the longitudinal wave-number $\eta$, the dimensionless critical load for the plate strip 
$p_{z, \mathrm{S}}$,~\EqRef{Eq-Plate-Critical-Load}, approximates the curves corresponding to $n=0$ in the dimensionless bifurcation load envelopes
shown in \FigureRef{Plot-Buckling} for thin shells. 

In order to capture this limit behaviour, a Taylor-series expansion in $\lambda_z$, truncated at the linear term about $\lambdaz = 1$, is introduced into the bifurcation condition~\eqref{Eq-Determinant}, with matrix $\tens{M}$ evaluated at $n=0$. The following critical stretch 
is obtained
\begin{equation} 
\label{lambdazapprox}
	\lambdaz\approx\frac{c_4(\tau,\nu)\,\eta^4+c_2(\tau,\nu)\,\eta^2+c_0(\tau,\nu)}{d_4(\tau,\nu)\,\eta^4+d_2(\tau,\nu)\,\eta^2+d_0(\tau,\nu)}\comma
\end{equation}
and neglecting the terms becoming inessential at large values of $\eta$, it can be further simplified to 
\begin{equation} 
\label{lambdazapproxsimp}
	\lambdaz \approx\frac{c_4(\tau,\nu)\,\eta^2+c_2(\tau,\nu)}{d_4(\tau,\nu)\,\eta^2+d_2(\tau,\nu)}\comma
\end{equation}
where
\begin{equation*}
	\begin{array}{l}
c_4=\tau ^3 \left(\tau ^2-12\right) \left[\tau ^2\left(51 \nu ^3-83 \nu ^2+12 \nu +17\right) +12 \left(17 \nu ^3-29 \nu ^2+4 \nu +7\right)\right] ,
\\[1.5mm]
c_2=12 \tau  \left[\nu  \tau ^4  \left(51 \nu ^3-55 \nu ^2-51 \nu +49\right)+12 \tau ^2\left(17 \nu ^4-24 \nu ^3-14 \nu ^2+22 \nu -3\right) -144 (1-\nu)^2
   (1+\nu)\right],
\\[1.5mm]
d_4=\tau ^3 \left(\tau ^2-12\right) \left[\tau ^2\left(51 \nu
   ^3-83 \nu ^2+9 \nu +20\right) +12 \left(17
   \nu ^3-29 \nu ^2+3 \nu +8\right)\right],
\\[1.5mm]
d_2=12 \tau  \left[\nu \tau ^4 \left(51 \nu ^3-55 \nu ^2-57 \nu
   +55\right) +12  \tau ^2 \left(17 \nu ^4-24 \nu ^3-16
   \nu ^2+24 \nu -3\right)-144 (1-\nu)^2 (1+\nu)\right].
\end{array}
\end{equation*}

At large longitudinal wave-numbers $\eta$, \EqRef{lambdazapproxsimp}, suitable for thin shells, which are characterized by small axial deformation before bifurcation, 
allows to compute the leading order approximation for the dimensionless pressure $\pz$ at bifurcation. 
An additional third-order series expansion around $\tau = 0$ leads to 
\begin{equation} 
\label{pz_asintotoP}
	\pz =\frac{\eta^2-2\nu}{12}\tau^2 + \bigO{\tau^4}.
\end{equation}
The above detailed procedure, based on an approximation of the bifurcation condition truncated at linear order in $\lambdaz$, was repeated assuming an expansion up to the third order, which led to a much more cumbersome equation with respect to~\EqRef{lambdazapproxsimp}, but yielded precisely the same result, ~\EqRef{pz_asintotoP}.

To allow a comparison with the plate strip solution, \EqRef{Eq-Plate-Critical-Load}, the asymptotic solution above, expressed in terms of current variables, as usual in bifurcation analysis, is to be restated in terms of reference variables,
thus an approximated explicit version of \EqRef{vars_update}$_2$ is sought. 
This equation is conveniently restated as $\eta^2 - \eta_0^2 \,\lambdar^2/\lambdaz^2=0$, 
which turns out to involve only $\eta,\, \eta_0,\, \nu,\, \tau_0$ upon introducing Eqs.~\eqref{Eq-Lambdar-Sol} and \eqref{lambdazapproxsimp} for $\lambdar$ and $\lambdaz$. 
Finally, the development of the latter condition into a Taylor series around $\tau_0=0$ up to the order 3, yields a bi-quadratic equation in $\eta$, whose solution gives the following approximated relationship, 
\begin{equation} 
\label{relaz_vars_iniz}
\eta^2 \approx \,\frac{2\,\eta_0^2\,\left(\nu\,\tau_0^2-3\,(1-\nu )\right)}{\eta_0^2\, \tau_0^2-6\,(1-\nu)}\comma
\end{equation}
such that, $\eta\to\eta_0$ as $\tau_0\to 0$. 
Noteworthy, \EqRef{relaz_vars_iniz} turns out to be valid for both the material models characterized by the strain energies $W_a$ and $W_b$.
Considering \EqRef{pz_asintotoP}, with the variables $\eta$ and $\tau$ replaced by $\eta_0$ and $\tau_0$ through Eqs.~\eqref{relaz_vars_iniz} and~\eqref{vars_update}$_1$, respectively, a final Taylor series expansion around $\tau_0=0$ up to the third order, leads to
\begin{equation} 
\label{pz_asintotoP_var_iniz}
	\pz =\frac{\eta_0^2-2\nu}{12}\tau_0^2+\bigO{\tau_0^4},
\end{equation}
therefore, for large longitudinal wave-numbers $\eta_0$, \EqRef{Eq-Plate-Critical-Load} is recovered asymptotically.

The dimensionless critical pressure for the plate strip, \EqRef{Eq-Plate-Critical-Load}, is superposed as a straight dashed line in the bi-logarithmic plot reported in \FigureRef{Plot-Buckling} for different values of $\tau_0$. The conclusion is that at large values of $\eta$, the plate theory provides a good approximation to the critical load of thin-walled  cylinders.
%
%
\subsection{Medium length cylinders: wrinkling}
\label{SubSec-Mid-Long-Shell}
As highlighted by \textcites{Timoshenko1961Student}, experiments show that thin cylindrical shells under compression usually buckle into short longitudinal waves, at a large longitudinal wave-number $\eta$. 
The bifurcation diagrams reported in Fig. \ref{Plot-Buckling} display an intermediate region where the buckling loads are almost independent of the values of both wave-numbers $n$ and $\eta$.
This region, for mildly long shells, corresponds 
to the so-called \lq wrinkling' (\textcite{Zhao2014}), for which  \textcite{Flugge1973} derived the critical load
\begin{equation}
\label{Eq-Shell-Critical-Load-Fluegge}
     p_{z,\mathrm{Fl\"ugge}} = \sqrt{\frac{1 - \nu^2}{3}} \, \tau_0.
\end{equation}
This classic solution can be rigorously recovered within the developed framework, by seeking the bifurcation condition as a minimum of the dimensionless axial pressure $p_z$ with respect to variable $\eta$. 
This corresponds to the stationarity of the bifurcation axial stretch evaluated for the mode $n=0$, \EqRef{lambdazapprox}, leading to 
five solutions. Among these, one is trivial, two are purely imaginary conjugated roots and two are real with opposite signs. From the latter pair, the positive real root is selected, 
\begin{equation}\label{Eq-Eta-Solution}
	\overbar{\eta} = 2\, \sqrt{\frac{\, e_1 +3\sqrt{e_2}}{e_3}}\comma
\end{equation}
where, assuming $\epsilon=17 \nu ^2-20 \nu +5$,
\begin{equation*}
   \begin{array}{l}
	e_1 = -3\, \nu ^2 \epsilon\, \tau ^3 \left(\tau ^2-12\right)\comma\\[2mm]
	e_2 = \left[144 \,\nu ^2\left(1-\nu ^2\right)^2 \tau-12\,\nu ^4\epsilon^2\, \tau ^3+\left[\nu\, \epsilon\, \tau ^2-12 \left(1-\nu^2\right)\right]^2 \log [(2-\tau)/(2+\tau)]\right]\tau ^3 \left(\tau ^2-12\right) \comma \\[2mm]
	e_3 =\left[\nu\, \epsilon\, \tau ^2-12 \left(1-\nu ^2\right)\right]\tau ^3 \left(\tau ^2-12\right).
   \end{array}
\end{equation*}
Equation \eqref{Eq-Eta-Solution} is now introduced into \EqRef{lambdazapprox} to evaluate the minimum  axial stretch for the mode $n=0$. The latter is finally used to compute the corresponding load, which is further expanded about $\tau=0$ to obtain, at first-order in $\tau$, (recalling~\EqRef{vars_update}$_1$) exactly the \Fluegge \EqRef{Eq-Shell-Critical-Load-Fluegge}.
%
%
\subsection{Slender cylinders: Euler rod buckling}
\label{SubSec-Euler-Buckling}
For a bar constrained with sliding clamps at both ends, assumed to be linearly elastic with \Young modulus $E$, the Euler buckling load  can be written in the form 
\begin{equation}
\label{Eq-Euler-Buckling-Classical-02}\pcomma
     N_{\mathrm{z,Euler}} =
     \frac{\pi ^3}{4} E\,a_0^2\, \alpha_0^2\,
   \tau_0 \left(4+\tau_0^2\right)\!,
\end{equation}
where $a_0=r_{e0}-t_0/2$, while $\alpha_0 = a_0/l_0$ represents the stubbiness ratio, in other words the inverse of the slenderness ratio in the reference configuration (being $\alpha = a/l$ its counterpart in the current configuration).

Euler buckling, affecting slender shells, characterized by a small stubbiness ratio  $\alpha_0$, corresponds to the anti-symmetric buckling mode with $m=1$, $n=1$.
The Euler formula, \EqRef{Eq-Euler-Buckling-Classical-02}, is recovered resorting to a perturbative technique \parencites{Golubitsky1985}{Simmonds1998}{Kokotovic1999}{Holmes2013} in the limit of vanishing longitudinal wave-number $\eta$. 
The approach followed by \cite{Goriely2008a} for incompressible materials is generalized by expanding both the radial and longitudinal stretches $\lambda_r (\alpha)$ and $\lambda_z (\alpha)$ 
in power series about $\alpha=0$ up to the order $M$,
\begin{equation}
\label{Eq-Asymptotic-Expansion-01}
    \lambdar(\alpha) = 1 + \sum_{i=1}^{M}{{\lambdar}_i \, \alpha^i} + \bigO{\alpha^{M+1}}, 
\quad 
\lambdaz(\alpha) = 1 + \sum_{i=1}^{M}{{\lambdaz}_i \, \alpha^i} + \bigO{\alpha^{M+1}}\comma 
\end{equation}
with coefficients ${\lambdar}_i$ and ${\lambdaz}_i$. 
The procedure is described here for the material with strain energy function  $W_a$ in~\EqRef{Eq-Def-Energy-PenceGou2015}, 
but parallel computations have been performed for the strain energy $W_b$, \EqRef{Eq-Def-Energy-PenceGou2015_separable}. 
The axisymmetry of pre-stress is enforced in~\EqRef{EqKr=0} in an approximate form, developing for convenience $\lambda_r^2\lambda_z T_{a_{rr}}$ in a Taylor-series about $\alpha=0$, with coefficients $k_j$ up to order $M$, 
\begin{equation}
\label{Krr-Asymptotic-Expansion-01}
    \lambda_r^2\lambda_z T_{a_{rr}} = \sum_{j=1}^{M}{k_j \, \alpha^j} + \bigO{\alpha^{M+1}}\point
\end{equation}
The relation between the coefficients ${\lambdar}_i$ and ${\lambdaz}_i$ is thus determined by requiring that the series in~\EqRef{Krr-Asymptotic-Expansion-01} vanishes at each order, so that the final approximation becomes
\begin{equation}
\label{Eq-Asymptotic-Expansion-r_planestress}
\lambdar(\alpha)= 1 +\sum_{i=1}^{M}{{\lambdar}_i({\lambdaz}_1,..,{\lambdaz}_M) \, \alpha^i} + \bigO{\alpha^{M+1}}. 
\end{equation}
To exemplify \EqRef{Eq-Asymptotic-Expansion-r_planestress},
when the order of approximation is $M=2$, the following coefficients are computed: 
${\lambdar}_1 = -\nu\lambda_{z1}$ and ${\lambdar}_2 = -\frac{1}{2} \, \nu  \left(\lambda_{z1}^2 \left(8 \nu ^2-11 \nu +2\right)+2 \lambda_{z2}\right)$.

The approximations,~Eqs.~\eqref{Eq-Asymptotic-Expansion-r_planestress} and \eqref{Eq-Asymptotic-Expansion-01}$_2$, are substituted in the bifurcation condition~\EqRef{Eq-Determinant}, with ${\tens{M}}$ computed for $m=n=1$; note that an infinitely slender cylinder buckles at vanishing load, namely, $\{\lambdaz, \alpha\} = \{1, 0\}$ represents a critical pair, therefore 
$\det{\tens{M}}_{|m=n=1}=0$ when $\alpha$ vanishes and thus $\lambdar=\lambdaz=1$. A further expansion into a Taylor series about $\alpha=0$ with coefficients $d_j$ up to order $N$, makes the buckling condition take the form
\begin{equation}
\label{det-Asymptotic-Expansion-01}
    \det{\tens{M}}_{|m=n=1}(\lambda_z(\alpha), \alpha, \tau, \nu) = \sum_{j=1}^{N}{d_j \, \alpha^j} + \bigO{\alpha^{N+1}}=0\point
\end{equation}
In order to satisfy this condition at each order, all coefficients $d_j$ are enforced to vanish. This leads to a system of linear equations for the unknown parameters ${\lambdaz}_i$. 
As the coefficients $d_j\, (j=1,2)$ turn out to vanish, 
$N=M+2$ is required to determine all the coefficients ${\lambdaz}_i$ ($i=1,..,M$) in~\EqRef{Eq-Asymptotic-Expansion-01}$_2$.

It turns out that ${\lambdaz}_i = 0$ for all odd values of the index $i=1,..,M$. Hence, the option $N=4$ is sufficient to provide the asymptotic expansion up to the third-order of $\lambdaz(\alpha)$,
\begin{equation}
\label{Eq-lambdaz-Euler-Asymptotic-z}
     \lambdaz(\alpha) = 1+\pi ^2\,\frac{  \left(\tau ^2+12\right)^2\nu ^2 - 36 \left(\tau ^2+4\right)}{288 \left(1-\nu ^2\right)}\; \alpha ^2 + \bigO{\alpha^4}\point
\end{equation}
The critical axial stretch in~\EqRef{Eq-lambdaz-Euler-Asymptotic-z} is defined with respect to the variables in the current configuration, and has to be related to the corresponding variables in the initial configuration to recover the critical load,  \EqRef{Eq-Euler-Buckling-Classical-02}. Therefore, the stubbiness ratio is expressed in terms of both current and initial variables as 
$\alpha=a/l=\alpha_0\lambda_r/\lambda_z$, so that
\begin{equation}
\label{stubbiness}\ppoint
     \alpha\lambda_z-\alpha_0\lambda_r=0.
\end{equation}
The asymptotic expansions~\eqref{Eq-Asymptotic-Expansion-r_planestress} and ~\eqref{Eq-Asymptotic-Expansion-01}$_{2}$ for $\lambdar$ and $\lambdaz$, respectively, plus 
a power series expansion about $\alpha_0$ for the function $\alpha$  (with coefficients $\alpha_k$), 
\begin{equation}
\label{alphavsalpha0}\ppoint
     \alpha =  \sum_{k=1}^{P}{ \alpha_k\, \alpha_0^k} + \bigO{\alpha_0^{P+1}}\comma
\end{equation}
are introduced into~\EqRef{stubbiness}. The obtained equation is solved at each order, thus obtaining the following expression (valid for $P=4$) 
\begin{equation}
\label{alphavsalpha0appr}
\alpha = \alpha_0 - \lambda_{z2}\,(1+\nu)\, \alpha_0^3+\bigO{\alpha_0^5}. 
\end{equation}

The longitudinal force resultant (positive when compressive) before bifurcation on the thin-walled tube can be finally computed as
\begin{equation}
\label{force}
N_{z} = -T_{zz}\pi(r_e^2-r_i^2)= -\lambda_r^2\, A_0\, T_{zz},
\end{equation}
so that inserting \EqRef{Eq-TensK-TensS-General}, and expanding the result in Taylor series about $\alpha_0 \to 0$ (slender columns), \EqRef{force} becomes
\begin{equation}
\label{Eq-Nzz-Euler-Asymptotic-N}
N_{z} = \frac{\pi ^3}{4} E\, a_0^2\, \alpha_0^2\, \tau_0 
\left[4 + \tau_0^2 - \frac{\nu ^2}{36\left(1-\nu^2\right)} \left(\tau_0^2-12\right) \tau_0^2 \right] + \bigO{\alpha_0^4}. 
\end{equation}

The buckling load  asymptotically derived from finite elasticity under the assumption of plane stress,~\EqRef{Eq-Nzz-Euler-Asymptotic-N}, can now be compared with the Euler buckling load,~\EqRef{Eq-Euler-Buckling-Classical-02}. 
It may be concluded that the two expressions for $N_z$ are identical at first-order in $\tau_0$, but differ at the third-order in $\tau_0$, because of a term 
depending on $\nu$, so that the coincidence up to the fourth-order occurs only when $\nu=0$. 
This little discrepancy remains very small for $\nu \in [0,0.5)$ when the dimensionless thickness $\tau_0$ is small, i.e. for thin shells. 
In fact, the relative difference $(N_{z}-N_{\mathrm{z,Euler}})/N_{\mathrm{z,Euler}}$ between the asymptotic approximation in~\EqRef{Eq-Nzz-Euler-Asymptotic-N} and the
usual formula for Euler's critical load,~\EqRef{Eq-Euler-Buckling-Classical-02}, is an increasing function of $\nu$ and $\tau_0$, attaining a maximum of 0.42\% as $r_e/r_i= 1.5$ ($\tau_0=0.4$); note that the latter is a value already far beyond the geometry of a thin shell. This is depicted for $\nu=0.3$ in \FigureRef{Plot-Buckling}, with the values of $\tau_0$ spanning within the large interval $[0,0.4]$. 

The asymptotic analysis has been repeated for the material with the strain energy function defined by~\EqRef{Eq-Def-Energy-PenceGou2015_separable}, which allows for the separation of the volumetric effects. This analysis has yield the same  asymptotic Euler buckling load up to order $\alpha_0^2$ given by~\EqRef{Eq-Nzz-Euler-Asymptotic-N}. 

It may be suggested that the \lq discrepancy factor' multiplied by $\nu$ may be a consequence of both the incremental plane stress assumption and the simplified kinematics underlying the two-dimensional approach presented here. 
In fact, for the material with strain energy function $W_a$ in~\EqRef{Eq-Def-Energy-PenceGou2015}, the incremental plane stress assumption becomes exact for $\nu=0$ and the radial stretch becomes unity, $\lambda_r=1$, as depicted in \FigureRef{fig:constitutive_laws}\,(a). For the material with strain energy  $W_b$,~\EqRef{Eq-Def-Energy-PenceGou2015_separable}, the incremental plane stress assumption has an order of accuracy  $\bigO{\alpha_0^2}$, and the radial stretch at bifurcation is approximated by the unity, $\lambda_r=1+\bigO{\alpha_0^4}$. 

It should be noticed that in \parencite{Goriely2008a}, the classical Euler buckling formula is exactly recovered up to the order $\alpha_0^2$ on the basis of a fully three-dimensional approach for an incompressible Mooney-Rivlin material.  
%
%
\section{3D bifurcation of a hollow (thick or not) cylinder}
\label{Sec-3D-Solution}
The fully three-dimensional solution for the bifurcation of a thick-walled cylinder made up of a hyperelastic material obeying the Pence-Gou model with strain energy $W_a$,~\EqRef{Eq-Def-Energy-PenceGou2015}, is derived in this Section, following the procedure outlined by \textcite{Chau1995} (see also \textcites{Chau1993}{Chau1998}) for a  class of materials characterized by an incremental constitutive law in the form
\begin{equation}
\label{jaujau}
	\begin{gathered}
		{\Jaumannderiv{T}}_{rr} = C_{11} \, \Dr + C_{12} \, \Dtheta + C_{13} \, \Dz \comma%
		\quad%
		{\Jaumannderiv{T}}_{\theta \theta} = C_{12} \, \Dr + C_{11} \, \Dtheta + C_{13} \, \Dz \comma\\%
		{\Jaumannderiv{T}}_{z z} = C_{31} \, (\Dr + \Dtheta)+ C_{33} \, \Dz \comma%
		\quad %
		{\Jaumannderiv{T}}_{r \theta} =  (C_{11} - C_{12}) \, \Drtheta \comma\quad
		{\Jaumannderiv{T}}_{\alpha z} =   2 \, C_{44} \, D_{\alpha z} \quad (\alpha = r, \theta);%
	\end{gathered}
\end{equation}
here the Zaremba-\Jaumann (or corotational) rate of the \Cauchy stress  $\Jaumannderiv{\tensT} = \tensdotS - (\tr{\tensD}) \, \tensT + \tensT \, \tensD - \tensW \, \tensT$ is adopted, as a function of $\tensD$.
The Pence-Gou model,~\EqRef{Eq-tensOldroydK-From-Energy-PenceGou}, fits the incremental form (\ref{jaujau}), when the coefficients $C_{ij}$ are defined as
\begin{equation}
	\label{Eq-Chau-Coefficients}
	\begin{gathered}
		C_{11} = %
		\kappa \, \lambdar^2 \,\lambdaz + %
		\mu \, \lambdaz^{-1}  \left(1+\lambdar^{-2}-2\,/3\, \lambdar^2\, \lambdaz^2\right) \comma%
		\quad %
		C_{33} = \kappa \, \lambdar^2 \, \lambdaz + %
			\mu\, \lambdar^{-2}\, \lambdaz^{-1} \left[1+\lambdaz^2\left(1-2\,/3\, \lambdar^4\right)\right]\comma\\
		C_{12} = C_{13} = C_{11} - 2 \mu \, \lambdaz^{-1} \comma\quad%
		C_{31} = C_{13} + \mu \, \lambdaz^{-1} \, \left( 1 - \lambdar^{-2} \, \lambdaz^2 \right) \comma\quad %
		C_{44} =\mu / 2 \, \left(\lambdar^{-2}\lambdaz+\lambdaz^{-1}\right)\point%
	\end{gathered}
\end{equation}
Note that the incremental moduli~\eqref{Eq-Chau-Coefficients} depend on the stretches $\lambdar$ and $\lambdaz$ in the pre-bifurcation state and on the constitutive parameters $\kappa$ and $\mu$.

The three incremental equilibrium equations for the linearized bifurcation problem can be decoupled through the introduction of the two potentials $\Phi(r,\theta,z)$ and $\Psi(r,\theta,z)$, such that 
\begin{equation}\label{Eq-Velocity-Potentials}
	\begin{dcases}
		\,\vr = \pardpard{\Phi}{r}{z} + r^{-1}\pardtheta{\Psi} \comma\\%
		\,\vtheta =r^{-1}\pardpard{\Phi}{\theta}{z}  - \pardr{\Psi} \comma\\%
		\,\vz = - \left[C_{11}\nabla_{1} \Phi + (1+s)\, C_{44}\, \parddz{\Phi} \right]/\left(C_{13} +  (1-s)\, C_{44}\right)\comma%
	\end{dcases}
\end{equation}%
where $s =  (\Tz-\Tr)/(2 \, C_{44})$ and $\nabla_{1} = r^{-1} \tfrac{\partial}{\partial r}\left(r\, \tfrac{\partial}{\partial r}\right) + r^{-2} \tfrac{\partial^2}{\partial \theta^2} $.
Equations~\eqref{Eq-Incremental-Equilibrium} can thus be written as
\begin{equation} \label{Eq-Incremental-Equilibrium-Velocity-Potentials-1}
	\begin{dcases}
		\,\left( \nabla_{1} - \nu_{1}^{\,2}  \frac{\partial^2}{\partial z^2} \right) \left( \nabla_{1} - \nu_{2}^{\,2}  \frac{\partial^2}{\partial z^2} \right)_{_{ }} \Phi = 0 \comma \\ 
		\,\left( \nabla_{1} + \nu_{3}^{\,2}  \frac{\partial^2}{\partial z^2} \right)^{^{ }} \Psi = 0 \comma
	\end{dcases}
\end{equation}
under the assumption that 
\begin{equation} \ppoint
	\nu_3^{\,2}  = 2\,  (1 + s)\, C_{44} /( C_{11} - C_{12})\comma
\end{equation}
with $\nu_1$ and $\nu_2$ representing the roots of the characteristic equation 
$A \, \nu_\alpha^4 + B \, \nu_\alpha^2 +C = 0$ ($\alpha=1,2$), with coefficients 
\begin{equation*}
	\begin{gathered}
		A =  (1 - s)\, C_{11} \, C_{44}\comma \,\,\,%
		B = C_{11} \, C_{33} - C_{13} \, C_{31} -  C_{44} \left[(1+s)\, C_{13} + (1-s)\,C_{31}  \right]\!\comma\,\,\,
		C =  (1 + s)\, C_{33} \, C_{44}\point 
	\end{gathered}
\end{equation*}
The regimes can be classified, according to the nature of the roots $\nu_1$ and $\nu_2$. 
The fulfillment of conditions  $B^2 - 4 \, A \, C >0$, $A \, C > 0$ and $B > 0$ 
define the elliptic-imaginary (EI) regime for the Pence-Gou material considered, where 
diffuse bifurcation modes are to be found  \parencite{Chau1995}.

The following representation for diffuse eigenmodal bifurcations are introduced via the above-introduced potentials
\begin{equation} \label{Eq-Velocity-Potentials-Specification}
	\begin{gathered}
		\Phi (r, \theta, z) = \phi(r) \, \cos{(n \, \theta)} \,\sin{(\eta \, z)}  \comma\\
		\Psi (r, \theta, z) = \psi(r) \, \sin{(n \, \theta)} \,\cos{(\eta \, z)} \comma
	\end{gathered}
\end{equation}
where $n$ and $\eta$ maintain the same definitions as in Eqs.~\eqref{Eq-Ansatz-Velocity-Bifurcation}. 
This choice of the potential functions, automatically satisfying the boundary conditions of free sliding along the faces $z=0$ and $z=l$, allows to write the equilibrium equations~\eqref{Eq-Incremental-Equilibrium-Velocity-Potentials-1} in the form
\begin{equation} \label{Eq-Incremental-Equilibrium-Velocity-Potentials-2}
	\begin{dcases}
		\,\left( \nabla_2 + \eta^2 \, \nu_1^{\,2} \right) \left( \nabla_2 + \eta^2 \, \nu_2^{\,2} \right) \phi = 0
		\\%
		\,\left( \nabla_2 - \eta^2 \, \nu_3^{\,2} \right) \psi = 0\comma
	\end{dcases}
\end{equation}
where $\nabla_2 =   r^{-1} \tfrac{\partial}{\partial r} (r \, \tfrac{\partial}{\partial r}) - n^2 \, r^{-2}$.
The general solutions to Eqs.~\eqref{Eq-Incremental-Equilibrium-Velocity-Potentials-2} are
\begin{equation}
\label{Velocity-Potentials-General-Solution}
	\begin{gathered}
		\phi(r) = b_1 \, H_{n}^{(1)}(\eta \, \nu_1 \, r) + b_2 \, H_{n}^{(1)}(\eta \, \nu_2 \, r) + b_3 \, H_{n}^{(2)}(\eta \, \nu_1 \, r) + b_4 \, H_{n}^{(2)}(\eta \, \nu_2 \, r) \comma
		\vspace{.1cm}\\
		\psi(r) = b_5 \, I_{n} (\eta \, \nu_3 \, r) + b_6 \, K_{n} (\eta \, \nu_3 \, r) \comma
	\end{gathered}
\end{equation}
where $H_{n}^{(1)}$ and $H_{n}^{(2)}$ represent the \Hankel functions of the first and second kind of order $n$, while $I_{n}$ and $K_{n}$ are the modified \Bessel functions of the first and second kind of order $n$, with coefficients $b_i$ being complex numbers. 

\begin{figure}[t!]
	\centering
	\includegraphics[width=0.85\textwidth]{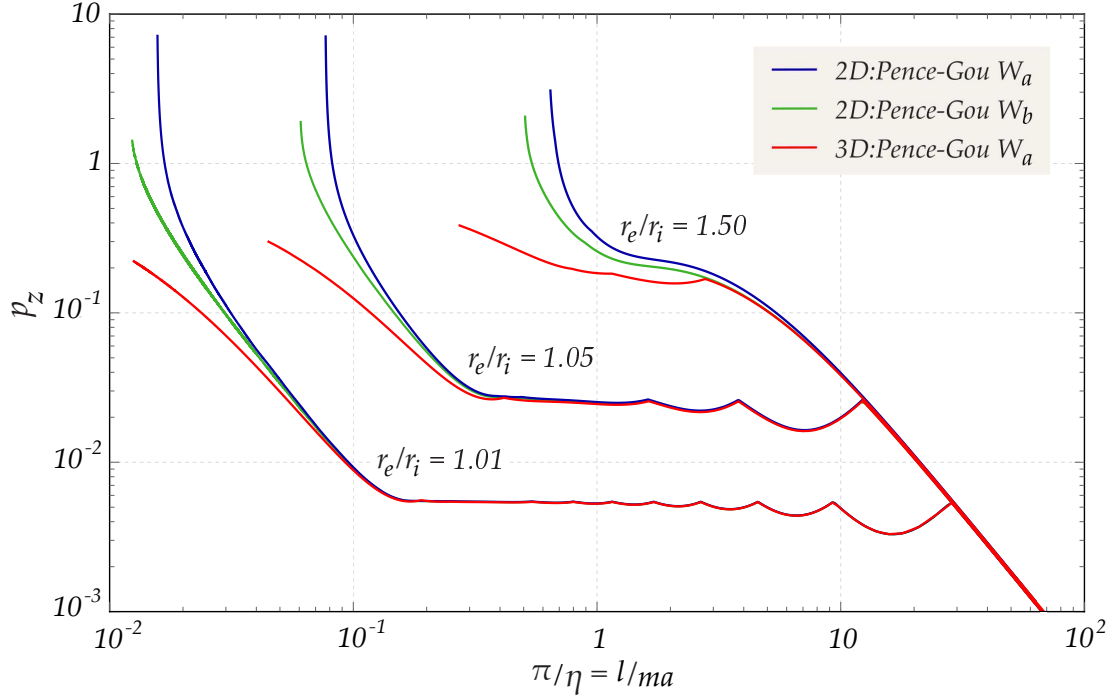}
	\caption{Comparison between the lower envelopes of
	dimensionless bifurcation loads $\pz$ for the buckling of cylindrical hollow cylinders, as a function of the longitudinal wave-number $\eta$ in a bi-logarithmic plot. Results from the thin-shell approximation, developed for Pence-Gou compressible materials with strain energies $W_a$ and $W_b$, are compared with the three-dimensional analysis performed for the Pence-Gou compressible materials with strain energy $W_a$; $\nu = 0.3$ is adopted.}
\label{fig:cfr}
\end{figure}
Enforcing the boundary conditions~\eqref{EqBC-dotS} of null tractions on both the inner and outer lateral surfaces of the pre-stressed cylinder, an eigenvalue problem in the form
$\bm{\mathcal{M}} \, \vec{b} = \vec{0}$ is obtained, with $\vec{b} = \{b_1, b_2, b_3, b_4, b_5, b_6\}^T$. Non-trivial solutions become possible when 
$\det{\bm{\mathcal{M}}} = 0$.
The latter condition only depends on the pre-bifurcation axial stretch $\lambdaz$, the dimensionless thickness of the shell $\tau$, the material parameter $\nu$, as well as the circumferential and longitudinal wave-numbers  $n$ and $\eta$. For a given set of parameters $\nu$, $\reoverri$, $n$ and $\eta$, the critical axial stretch can be found numerically.

A comparison is reported in \FigureRef{fig:cfr} between the critical envelopes evaluated on the basis of the 3D approach and the thin-shell approximation. 
The 3D approach has been developed for a compressible Pence-Gou material with the strain energy $W_a$, while results for the thin shell approximation are reported for both strain energies $W_a$ and $W_b$.
Geometry of the cylinder varies between very thin-, thin- and medium-walled. 

It should be noted that the three-dimensional approach fully 
captures the nearly constant branch of the curve, corresponding to the asymptotic load derived by \Fluegge for medium length cylinders, \EqRef{Eq-Shell-Critical-Load-Fluegge}.

The accuracy of the thin-shell approximation is evident from the comparison with the three-dimensional solution described in the present Section.
In particular, the critical modes characterized by small longitudinal wave-numbers $\eta$ are neither altered by the chosen approach, nor by the constitutive model adopted, so that the curves reported in \FigureRef{fig:cfr} are  almost coincident within the most important part of the buckling landscape. On the contrary, for modes with small circumferential wave-numbers (in particular for $n=0$, critical for large longitudinal wave-numbers $\eta$, Fig. \ref{Plot-Buckling})  a noticeable difference between the curves becomes appreciable, becoming more evident when the thickness of the shell increases. This discrepancy is due to the fact that a surface bifurcation is approached and thus the thin-walled solution is no longer valid.

Of course the thin-shell approximation is much more efficient from the computational point of view (with CPU times for the single evaluation of a critical pair $\{\lambda_r,\eta\}$ according to the approximated approach getting as low as 1/300 of the times for the fully three-dimensional approach), however, the adopted hypothesis of incremental plane stress becomes unreliable as the shell thickness grows.
%
%
\section{Conclusions}

A complete re-derivation has been presented for the bifurcation  
of axially compressed thin-walled cylinders. The most important aspect of the new formulation is the independence from the constitutive equation used in the original formulation by \Fluegge, which does not stem 
from any strain potential and is now replaced by a generic nonlinear law of elasticity. 
Using two different hyperelastic constitutive laws, we have  
rigorously confirmed the results by \Fluegge, together with several limit formulae (for surface instability, wrinkling, and Euler rod buckling). 
The outlined approach allows now the precise and computationally efficient analysis of the bifurcation landscape for a thin-walled cylinder adopting all other related formulae for any constitutive law characterizing soft materials. 

\vspace{5mm}
{\it Acknowledgments} 
The authors acknowledge  
financial support from 
ERC-ADG-2021-101052956-BEYOND.
%
%
%
%
\newline
\printbibliography
\end{document}